\documentclass[12pt]{iopart} 
\usepackage{epsfig,amssymb}

\begin{document}

\title[]{The first WIMPy halos}

\author{Anne M.~Green\dag\ddag\S, Stefan Hofmann\S and Dominik
J.~Schwarz$\|$}

\address{\dag\ Physics and Astronomy, University of Sheffield, Sheffield
    S3 7RH, UK}
\address{\ddag\ Astronomy Centre, Department of Physics and
  Astronomy, University of Sussex, Falmer, Brighton, BN1 9QH, UK} 
\address{\S Physics Department, Stockholm
  University, 106 91 Stockholm, Sweden} 
\address{$\|$ Fakult\"at f\"ur Physik, Universit\"at Bielefeld, 
  Postfach 100131, 33501 Bielefeld, Germany}
      
\eads{\mailto{a.m.green@sheffield.ac.uk}, \mailto{stehof@physto.se},
\mailto{dschwarz@physik.uni-bielefeld.de}}

\begin{abstract}

Dark matter direct and indirect detection signals depend crucially on
the dark matter distribution. While the formation of large scale
structure is independent of the nature of the cold dark matter (CDM),
the fate of inhomogeneities on sub-galactic scales, and hence the
present day CDM distribution on these scales, depends on the
micro-physics of the CDM particles. We study the density contrast of
Weakly Interacting Massive Particles (WIMPs) on sub-galactic scales.
We calculate the damping of the primordial power spectrum due to
collisional damping and free-streaming of WIMPy CDM and show that 
free-streaming leads to a CDM power spectrum with a sharp cut-off at about
$10^{-6} M_\odot$. We also calculate the transfer function for the
growth of the inhomogeneities in the linear regime, taking into
account the suppression in the growth of the CDM density contrast
after matter-radiation equality due to baryons and show that our
analytic results are in good agreement with numerical calculations.
Combining the transfer function with the damping of the primordial
fluctuations we produce a WMAP normalized primordial CDM power
spectrum, which can serve as an input for high resolution CDM
simulations. We find that the smallest inhomogeneities typically have
co-moving radius of about $1$ pc and enter the non-linear regime
at a redshift of $60 \pm 20$. We study the effect of scale dependence
of the primordial power spectrum on these numbers and also use the
spherical collapse model to make simple estimates of the properties of
the first generation of WIMP halos to form. We find that the very first
WIMPy halos may have a significant impact on indirect dark matter
searches. 
\end{abstract}


\begin{flushleft} {\bf Keywords}: dark matter, cosmological perturbation theory
\end{flushleft}


\maketitle

\section{Introduction}
\label{intro}

Analysis of the anisotropies in the cosmic microwave background (CMB)
radiation~\cite{sperg} finds that the relative matter density
$\Omega_{{\rm m}} = 0.29 \pm 0.07$ is significantly larger than the
relative baryon density $\Omega_{{\rm b}} = 0.047 \pm 0.006$. This is
consistent with the observed abundances of light elements and
primordial nucleosynthesis (see e.g. Reference~\cite{nucleo}) and the
power spectrum found from galaxy red-shift surveys~\cite{galaxy}, and
indicates that the Universe contains a significant amount of
non-baryonic cold dark matter (CDM). The identification of the nature
of the CDM particles is a major outstanding issue for cosmology (and
would provide reassuring confirmation of the CDM cosmological
paradigm).

There are various CDM candidates
(for a recent review see~\cite{gondolorev}), the most well studied of
which are Weakly Interacting Massive Particles (WIMPs) and axions.
WIMPs are a particularly attractive CDM candidate, since a stable
relic from the electroweak scale generically has an interesting
present day density, $\Omega_{\rm wimp} \sim {\cal O}(1)$~\cite{dimop}.
There are a large number of ongoing experiments attempting to detect
WIMPs directly in the lab~\cite{direct} or indirectly via their
annihilation products (gamma-rays, antiprotons and neutrinos)
~\cite{indirect}.

The signals expected in dark matter detection experiments depend in
most cases on the distribution of the dark matter. Direct
detection experiments probe the dark
matter distribution on sub-milli-pc scales~\cite{SS,submilli} while
indirect detection signals are strongest from the
highest density regions of the Milky Way (see
e.g. References~\cite{SS,clump1}) and the extra-galactic gamma-ray
signal depends on the clumpiness of the WIMP
distribution~\cite{clump2}.  Reliable predictions for the expected
signals therefore require an understanding of the clumpiness of dark
matter on small (sub-galactic) scales.  The density perturbations on
very small scales, and hence the properties of the first generation of
structures to form, depend on the microphysics of the CDM and the
present day density distribution may retain traces of these first
structures.

Two of the present authors showed that the microphysics of WIMPs
(specifically collisional damping due to interactions with the
radiation component and free-streaming) lead to a fundamental small
scale cut-off in the WIMP density perturbation power
spectrum~\cite{hss,shs}. Subsequently we presented the small scale
WIMP density, velocity and potential perturbations and estimated the
properties of the first generation of WIMP halos to form for the case
of the WIMP being a bino~\cite{ghs}. Meanwhile Berenzinsky et
al.~\cite{berenzinsky} studied the survival probability of these first
halos analytically. More recently Diemand et al.~\cite{dms} have
carried out high resolution simulations of the first WIMP halos to
form and there has been (inconclusive) discussion about whether these
halos will be disrupted by interactions with stars~\cite{disrupt}.
Shortly after the first version of the present work became
available on arXiv.org, Loeb and Zaldarriaga published a numerical
calculation of the cut-off scale \cite{LZ}. Their work is an
improvement on the analytic treatment in the present paper, although
the estimates for the cut-off mass scales agree up to a factor of
order unity (the difference between their $10^{-4} M_\odot$ and our
$10^{-6} M_\odot$ stems largely from different assumptions for the
kinetic decoupling temperature -- $10$ MeV in \cite{LZ} versus $30$ MeV
for our benchmark model in \cite{ghs}, which is a somewhat more
realistic value for the lightest neutralino).

In this paper we present the detailed calculations behind the results
presented in our earlier letter~\cite{ghs} generalised to generic
WIMPs, including a new more intuitive collisional damping calculation
using the Navier-Stokes equation.  We also compare our analytic
transfer function with numerical calculations and study the effects of
a scale dependent primordial power spectrum. In Section~\ref{wimps} we
estimate the temperature scales related to chemical and kinetic
decoupling for generic WIMPs.  In Section~\ref{pert} we define the
fluid perturbation variables and present the equations necessary for
our subsequent calculations and solve the evolution of the matter and
radiation perturbations during the radiation dominated regime. In
Sections~\ref{kdsec} and~\ref{fs} we calculate the damping of the WIMP
density contrast due, respectively, to collisional damping (as a
result of elastic interactions with other species) and free-streaming.
We then (Section~\ref{evol}) calculate the transfer functions, which
encode the evolution of the perturbations, by solving the evolution
equations around matter-radiation equality and matching these
solutions to the sub-horizon limits of the radiation domination
solutions found in Section~\ref{pert}, and compare our expressions
with the output of the COSMICS package.  This allows us to calculate
the power spectrum for the WIMP perturbations on sub-galactic scales
shortly after matter-radiation equality (Section~\ref{ps}).  We
consider the case of a scale-invariant primordial power spectrum and
discuss the modifications due to tilt and running of the spectral
index for single-field inflationary models. Finally in
Section~\ref{results} we estimate the epoch at which typical-sized
inhomogeneities go non-linear for a range of benchmark WIMP models and
examine the effects of scale dependence of the primordial power
spectrum on our results. To conclude we also estimate the epoch when
the very first (rare) inhomogeneities go non-linear and estimate their
size after collapse and speculate about their fate and relevance for
dark matter searches.

\section{WIMPs}
\label{wimps}
Weakly interacting massive particles are generic candidates for CDM. 
The reason is that they are quite natural in extensions of the standard
model of particle physics, the most popular example being supersymmetry. 
In supersymmetry models every standard model particle has a supersymmetric
partner and in most models there is a conserved quantum number
(R-parity), which makes the lightest supersymmetric particle
stable. Supersymmetry models have a large number of free parameters,
however in most models the lightest supersymmetric particle is the
lightest neutralino (which is a mix of the supersymmetric partners of
the photon, the $Z$ and the Higgs bosons) and an excellent CDM
candidate (see e.g. Reference \cite{jkg}). We have focused our
attention on the neutralino (in particular a bino-like neutralino) in
our previous work \cite{ghs}. Here we wish to dwell on the model-independent
aspects of WIMP astrophysics. Our treatment is valid given three
assumptions hold true: we assume that there is no WIMP anti-WIMP
asymmetry in the universe and we assume that WIMPs have been in
chemical and thermal equilibrium with the radiation component in the
early (hot) Universe (thus Wimpzillas are different from what we call
generic WIMPs). The third assumption is that the elastic cross sections
are dominated by $Z^0$ exchange. This assumption is wrong for bino-like
neutralinos, but as our previous studies have shown, the final result is
not very sensitive to this detail. For simplicity, we also assume that
there is only one CDM component, namely WIMPs. 

In the early universe WIMPs can be treated as an ideal Bose or Fermi
gas characterised by $g$ internal degrees of freedom, the WIMP mass $m$, 
chemical potential $\mu$ and temperature $T$. Unless the WIMPs have a net 
non-zero quantum number (in which case a mechanism is required to generate a
WIMP anti-WIMP asymmetry) the chemical potential is negligible and,
independent of the spin statistics, for $T \ll m$ and $\mu \ll m$ 
the distribution function takes the Boltzmann form and the relative relic 
abundance of WIMPs is given by:
\begin{equation}
\label{om}
\Omega_{\rm wimp}=
\frac{2}{3} \left(\frac{2^3}{\pi}\right)^{1/2} \!\! \left.
\frac{g_{\rm s} T^3}{(m_{\rm Pl} H)^2}\right|_0
\frac{g m}{g_{\rm s}|_{\rm cd}}
\left(x_{\rm cd}\right)^{3/2} \exp\left(-x_{\rm cd}\right) 
\cosh\left(x_{\rm cd}\frac{\mu_{\rm cd}}{m}\right) \,.
\end{equation} 
Here $m_{\rm Pl}$ denotes the Planck mass, $H$ the Hubble rate, $g_s$ 
the number of
relativistic degrees of freedom contributing to the total entropy of the
radiation fluid and
$x\equiv m/T$. The subscripts `0' and `cd' refer to the present day
and the epoch at which WIMP anti-WIMP annihilation ceased and the
WIMPs chemically decoupled, respectively. In the following we assume 
$\mu_{\rm cd} = 0$. 

Equation (\ref{om}) can be solved iteratively for $x_{\rm cd}$:
\begin{eqnarray}
x_{\rm cd}
&\approx&
x_{\rm cd}^{(0)} + \frac{3}{2}\;  {\rm ln} \; x_{\rm cd}^{(0)}
\; ,\nonumber \\
x_{\rm cd}^{(0)}
&\approx&
23 + {\rm ln}\left(\frac{m}{100 \; {\rm GeV}}\right) - 
{\rm ln}\left(\Omega_{\rm wimp} h^2\right) + {\rm ln}\left(g\right) 
+ {\rm ln}\left(\frac{g_{\rm s}|_0}{g_{\rm s}|_{\rm cd}}\right) 
\; .
\end{eqnarray}

After chemical decoupling, $T \lesssim T_{\rm cd}$ the total number of
WIMPs remains constant and they are kept in local thermal equilibrium
by elastic scattering processes with relativistic particles: ${\rm
WIMP} + {\rm L} \longleftrightarrow {\rm WIMP} + {\rm L}$. As the
Universe expands the WIMP density decreases, the elastic scattering
rate decreases and the WIMPs kinetically decouple. The characteristic
time scale between elastic scatterings is $\tau_{\rm f}\equiv
1/\Gamma_{\rm el}$, where $\Gamma_{\rm el}$ denotes the elastic
scattering rate.  The average momentum exchanged per collision is
small, of order $T$ \cite{ssw,hss}, and the number of elastic
scatterings needed to keep the WIMPs in local thermal equilibrium,
$N$, is large: $N \approx m/T \gg 1$. The relaxation timescale,
$\tau_{\rm r}$, which characterizes the time at which the WIMP
kinetically decouple, is given by $\tau_{\rm r}\equiv N \tau_{\rm f}$
and is significantly larger than the elastic scattering timescale.

The elastic scattering rate is given by
\begin{equation}
\label{es}
\Gamma_{\rm el}\equiv 
\sum\limits_{{\rm L}\in {\rm SM}} \langle v \sigma_{\rm el}\rangle n_{\rm L}
\; ,
\end{equation}
where $\sigma_{\rm el}$ is the total cross section for elastic
scatterings of WIMPs and relativistic Standard Model fermions, $n_{\rm
L}$ the number density of relativistic particles of species $L$, which
are assumed to be in local thermal equilibrium and $v\approx 1$ in this case.  
The thermal
average of $\sigma_{\rm el}$ can be written as $\langle \sigma_{\rm
el}\rangle = \sigma^{\rm el}_0 (T/m)^{1+l}$, where $\sigma^{\rm el}_0
\approx (G_{\rm F} m_{\rm W}^{\; 2})^2 m^2/m_{\rm Z}^{\; 4}$ sets the
scale for the cross section and $l$ parametrises the temperature
dependence.  Here, $m_{\rm W}$ denotes the mass of the charged gauge
bosons in the electroweak interaction, $m_{\rm Z}$ is the mass of the
neutral gauge boson and $G_{\rm F}$ is Fermi's coupling constant.  In
the Standard Model, elastic scattering between a light fermion and a
heavy fermion is mediated by ${\rm Z}^0$ exchange and $l=0$. Other
channels may occur however. In supersymmetric extensions of the
Standard Model, where the lightest neutralino is a WIMP candidate,
sfermion exchange occurs (and if the neutralino is a gaugino,
${\rm Z}^0$ exchange is suppressed), in which case $l=1$.

Kinetic decoupling of WIMPs happens at a temperature $T_{\rm kd}$,
defined by $\tau_{\rm r} (T_{\rm kd}) = H^{-1}(T_{\rm kd})$.
Solving this equation for $x_{\rm kd}\equiv m/T_{\rm kd}$, we find 
\begin{eqnarray}
x_{\rm kd}
&&\approx 
\left[\frac{\zeta(3)}{\pi^2} 
\left(\frac{90 g_\epsilon\mid_{\rm kd}}{8\pi^3}\right)^{1/2} m_{\rm Pl} \; 
m \; \sigma_0^{\rm el} (m)\right]^\frac{1}{3+l}
\,, \nonumber \\
&& \approx
\left[7\cdot 10^{13} g_\epsilon^{1/2} \mid_{\rm kd} \; 
\left(\frac{m}{100 \; {\rm GeV}}\right)^3 
\right]^\frac{1}{3+l}
\; ,
\label{xkd}
\end{eqnarray}  
where $g_\epsilon^{1/2}$ is the number of degrees of freedom contributing
to the energy density.  For $l= 0$ the kinetic decoupling temperature does
not depend on the WIMP mass and is given by $T_{\rm kd} \approx 2.4 \;
g_\epsilon^{-1/6}\mid_{\rm kd}$ MeV.  For $l=1$, $T_{\rm kd} \approx 34.2
\; g_\epsilon^{-1/8}\mid_{\rm kd} (m/100 {\rm GeV})^{1/4}$ MeV.

\begin{figure} 
\begin{center} 
\epsfxsize=6.in 
\epsfbox{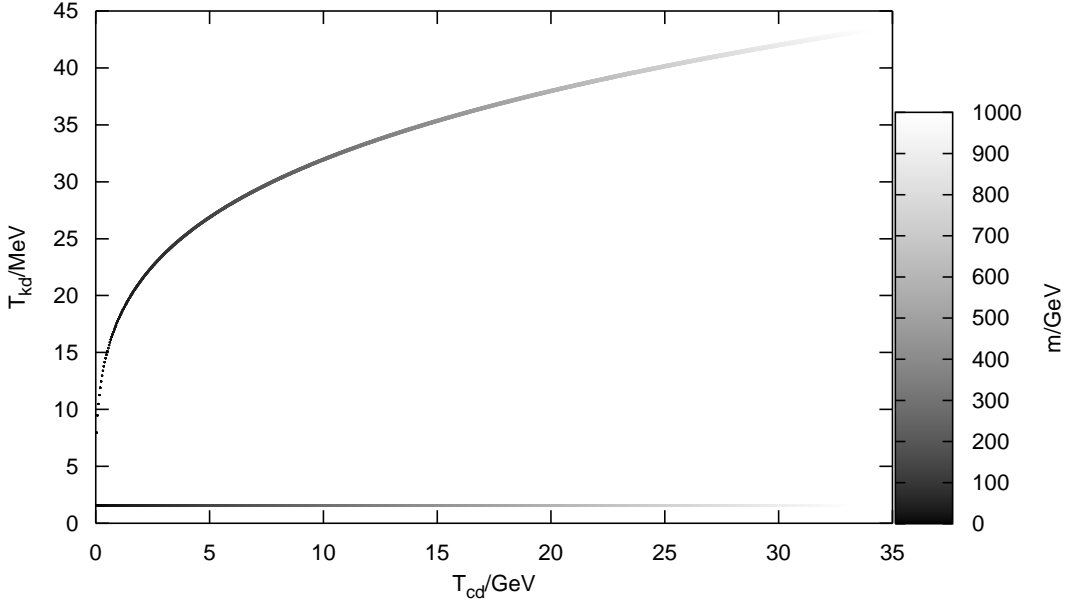}
\end{center} 
\begin{center}
\caption{The dependence of the WIMP chemical and kinetic decoupling
temperatures, $T_{\rm cd}$ and $T_{kd}$, on the WIMP mass (indicated by
the grey scale) for WIMPs with $\omega_{\rm wimp} = 0.076 - 0.156$.
The upper and lower bands are for l=1 (Majorana) and 0 (Dirac) respectively.
\label{TkdTcd}} 
\end{center}
\end{figure}

In figure~\ref{TkdTcd} we plot the dependence of the WIMP chemical and
kinetic decoupling temperatures on the WIMP mass, for WIMPs with relic
densities corresponding to the WMAP measurement of the CDM
density (i.e. assuming that the cold dark matter is entirely in the
form of WIMPs) $\omega_{\rm cdm}= \Omega_{\rm cdm}
h^2=0.076 - 0.156$, consistent with the 2-sigma error of WMAP \cite{sperg}. 

While chemical decoupling happens at a temperature of around $10$ GeV,
kinetic decoupling is delayed by the large entropy of the hot Universe
and takes place at around $10$ MeV. This is generic for any WIMP
that is of cosmological relevance today.

\section{Cosmological perturbations of fluids}
\label{pert}

Fluctuations in the energy density of radiation and matter come together 
with fluctuations of the scalar metric potentials. In the conformal 
Newtonian (or longitudinal or zero shear) gauge, perturbations of an 
isotropic and homogeneous, spatially flat metric are characterised by 
two scalar potentials $\phi$ and $\psi$ which appear in the line element 
as (e.g. Reference \cite{mfb})
\begin{equation}
{\rm d}s^2
=
a^2(\eta)\left[-\left(1+2\phi\right){\rm d}\eta^2
+\left(1-2\psi\right) \delta_{ij} \; {\rm d}x^{i}{\rm d}x^{j}
\right]
\; ,
\end{equation}
where $a$ denotes the scale factor and $\eta$ conformal time. We work in 
this gauge since all sub-horizon quantities can be interpreted in terms of
Newtonian physics, in particular the scalar potential $\phi$ is identical 
to the gravitational potential in the Newtonian limit. This is in contrast 
to the synchronous gauge where the Newtonian gravitational potential is 
gauged to zero. Furthermore, in the conformal Newtonian gauge there is no 
complicating residual gauge degree of freedom.

For an isotropic fluid with energy density $\epsilon$, pressure $P$ and 
four-velocity $u^\mu$, the components of the energy-momentum tensor are 
given by
\begin{equation}
\label{emti}
T^\mu_{\; \nu} = (\epsilon +P) u^\mu u_\nu + P \delta^{\mu}_{\nu} \; .
\end{equation}
Perturbations in the energy density and pressure are denoted by 
$\delta\epsilon$ and $\delta P$ respectively, and we treat the (small) fluid 
peculiar velocity $v_i$ as a linear velocity perturbation. The peculiar 
velocity is related to the spatial component of the four-velocity by
$u^i = v^i/a$. The components of the perturbed energy-momentum tensor
are given to first order in the perturbations by
\begin{equation}
\delta T^0_{\; 0} = - \delta\epsilon \; , \quad
\delta T^0_{\; j} = \frac 1a \left(\epsilon + P\right)v_j \; , \quad
\delta T^i_{\; j} = \delta^i_j\; \delta P + \Pi^i_{\; j}  
\end{equation}
Here, $\Pi^i_{\; j}\equiv T^i_{\; j} - \delta^i_j T^k_{\; k}/3$ is the
traceless anisotropic pressure. Quantum fluctuations during inflation
do not seed anisotropic pressure, however once neutrinos decouple from
the photon-lepton fluid at about $T \sim 1$ MeV, they build up
anisotropic pressure, which has to be taken into account to obtain the
correct normalisation of the power spectra. We will include this
effect below, but neglect the anisotropic stress of the neutrinos for
all other aspects in this paper. 

In a spatially flat background it is convenient to consider Fourier
modes with wavenumber ${\bf k}$. For linear perturbations these modes
are decoupled from each other and, due to the isotropy and homogeneity
of the background, all modes with the same $k = |{\bf k}|$ obey
identical mode equations.

We define dimensionless scalars to characterise the fluctuations of each
fluid
$\Delta_{\rm a}$
\begin{equation}
\Delta_{\rm a} \equiv \frac{\delta\epsilon_{\rm a}}
{(\epsilon + P)_{\rm a}} \; ,
\end{equation}
and $v_{\rm a}$, the modulus of the peculiar velocity,
\begin{equation}
v_{\rm a} \equiv -\imath {\bf \hat{k}} \cdot {\bf v}_{\rm a} \; ,
\end{equation}
with the index index ``a'' denoting the type of fluid (e.g. radiation 
$P_{\rm r}=\epsilon_{\rm r}/3$ or CDM $P_{\rm cdm}=0$). 

It is also useful to define the total energy density contrast $\Delta$ and
peculiar velocity modulus $v$:
\begin{equation}
\Delta
= \sum\limits_{\rm a}
\frac{\left(\epsilon + P\right)_{\rm a}}{\epsilon + P}
\; \Delta_{\rm a}
\; ,\quad
v =
\sum\limits_{\rm a}
\frac{\left(\epsilon + P\right)_{\rm a}}{\epsilon + P}
\; v_{\rm a}
\; ,
\end{equation}
which act as sources of the energy and momentum constraints
respectively.  For fluids, the pressure can be related to the energy
density and the entropy by an equation of state $P_{\rm a} = P_{\rm
a}(\epsilon_{\rm a}, S_{\rm a})$. We need only consider the
situation of adiabatic and isentropic evolution of each fluid, thus
$\delta S_{\rm a} = 0$, and hence
\begin{equation}
\delta\, P_{\rm a} = c_a^2 \delta \epsilon_{\rm a}\; , \quad 
c_{\rm a}^2 = 
\left(\frac{\partial P_{\rm a}}{\partial\epsilon_{\rm a}}\right)_S\; ,
\end{equation}  
where $c_{\rm a}$ is the adiabatic sound speed of fluid ``a''. 

Each of the fluids individually (if they are noninteracting and
dissipationless) obeys the general relativistic continuity and Euler
equations. These equations read (here we have used $c_{\rm a}^2 =
P_{\rm a}'/\epsilon_{\rm a}'$, which is valid for linear
perturbations only):
\begin{eqnarray}
\label{f1}
\Delta_{\rm a}^{\; \prime} &=& k v_{\rm a} + 3 \psi^\prime
\; , \\
\label{f2}
v_{\rm a}^{\; \prime} + (1 - 3 c^2_{\rm a}) {\cal H} v_{\rm a}
&=&
- c^2_{\rm a} k \Delta_{\rm a} 
- k \phi + \frac 23 k \pi_{\rm a}
\; ,
\end{eqnarray}
where the primes denote derivatives with respect to conformal time
$\eta$ and ${\cal H} \equiv a'/a$. Anisotropic pressure is generated
by $\pi$ which is defined by $\Pi^i_j/(\epsilon + P) = (\delta^i_j/3 -
\hat{k}^i \hat{k}_j) \pi$.

Any number of perfect fluids can be described 
by one effective imperfect fluid. The effective 
dimensionless entropy perturbation is
\begin{equation}
\label{sourceS}
{\cal S} \equiv {\delta P - c_s^2 \delta \epsilon\over \epsilon+P} =
\sum_a c_{\rm a}^{\; 2} \; \frac{(\epsilon + P)_{\rm a}}{\epsilon + P}\;
\left( \Delta_{\rm a} - \Delta\right) \; ,
\end{equation}
where the effective adiabatic sound speed is given by
\begin{equation}
c_{\rm s}^{\; 2} = \sum_a c_{\rm a}^2 
\frac{(\epsilon + P)_{\rm a}}{\epsilon + P}\; .
\end{equation}
For a Universe containing CDM and radiation 
fluids we find
\begin{equation}
{\cal S} = {y \over 4 + 3 y} \left( \Delta_{\rm r} - \Delta_{\rm cdm}
           \right)\; ,
\quad c_{\rm s}^2 = \frac 13 \; {1\over 1 + 3 y/4}\; ,
\end{equation}
where $y \equiv a/a_{\rm eq}$ is the scale factor normalised at 
matter-radiation equality. 

We can now define isentropic initial conditions by demanding that
${\cal S} = 0$ and ${\cal S}'= 0$. The first condition gives
$\Delta_{\rm cdm} = \Delta_{\rm r}$ and the second, making use of the
continuity equation (\ref{f2}), gives $v_{\rm cdm} = v_{\rm r}$
additionally. The generalisation to an arbitrary number of fluids is
straightforward.  Below we restrict our attention to these
(isentropic) initial conditions. 

In order to close the set of equations we also need the 
Einstein equations; the background equation 
\begin{equation}
{\cal H}^2 -{\cal H}' = 4\pi G a^2 (\epsilon + P) \,,
\end{equation}
and the energy and momentum constraints, which have $\Delta$ and
$v$ as source terms respectively,
\begin{eqnarray}
\label{e1}
- k^2\psi - 3{\cal H}\psi^\prime - 3{\cal H}^2\phi
&=& \left({\cal H}^2-{\cal H}^\prime\right) \Delta
\; , \\
\label{e2}
- k\left(\psi^\prime + {\cal H}\phi\right)
&=& \left({\cal H}^2-{\cal H}^\prime\right) v 
\; .
\end{eqnarray} 
Combining the spatial
trace of the Einstein equations and the energy constraint one can
write down an equation which has ${\cal S}$ as a source (see
e.g. Reference \cite{mfb}), but we do not need it here. The spatial
off-diagonal Einstein equation couples the anisotropic pressure to the
metric potentials, such that
\begin{equation}
k^2 (\psi - \phi) = 2 ({\cal H}^2 - {\cal H}') \pi \; .
\end{equation}
For isotropic fluids, $\pi = 0$ and hence $\psi = \phi$. 
In general the difference between $\psi$ and $\phi$ can
be neglected on subhorizon scales ($k \gg {\cal H}$). 

To normalise the spectra of cosmological perturbations, we have to
connect solutions in the radiation, matter and dark energy dominated
epochs with the primordial fluctuations that are produced during the
epoch of cosmological inflation. To do that it is most convenient to
introduce a quantity that is independent of the choice of the
hypersurface of constant time and 
is constant for modes in the superhorizon regime ($k
\ll {\cal H}$). Following Bardeen \cite{bard} we define
\begin{equation}
\zeta = \Delta/3 - \psi\; ,
\end{equation}
which is the curvature perturbation on uniform density hypersurfaces
or the density contrast on uniform curvature hypersurfaces.  We see
immediately from the continuity equations (\ref{f1}), that this
quantity could be defined for each fluid or for one effective fluid
and that it is constant for regular solutions on superhorizon scales
(assuming that there are no entropy perturbations) \cite{zeta}. In the 
following we normalise the power spectrum of $\zeta$ to the value measured
by WMAP, where this variable is denoted $-{\cal R}$.

During radiation domination ($\epsilon_{\rm r} \gg \epsilon_{\rm cdm}$),
the perturbation equations (\ref{f1}) and (\ref{f2}) can be solved
exactly for the cold dark matter and isotropic radiation 
($\pi=0 \Rightarrow \phi =
\psi$) fluids, on all scales
\cite{ssw}. In the conformal Newtonian gauge we find for the dominant growing
mode 
\begin{eqnarray}
\label{gradex}
\phi^{^{\rm rad}}(\kappa)
&=&
- 2 \zeta_0 \frac{j_1(\kappa)}{\kappa}
\; ,\nonumber\\
\Delta_{\rm r}^{^{\rm rad}}(\kappa)
&=&
3 \zeta_0 \left[
j_0(\kappa)-2\frac{j_1(\kappa)}{\kappa}+\kappa j_1(\kappa)\right]
\; , \nonumber \\
v_{\rm r}^{^{\rm rad}}(\kappa)
&=&
\sqrt{3} \zeta_0
\left[\kappa j_0(\kappa)-2 j_1(\kappa)\right]
\; , \\
\label{gmatex}
\Delta_{\rm cdm}^{^{\rm rad}}(\kappa)
&=&
6 \zeta_0 \left[
{\rm ln}(\kappa) + j_0(\kappa) - \frac{j_1(\kappa)}{\kappa} - {\rm Ci}(\kappa) 
+ \gamma_{\rm E}-\frac{1}{2}
\right]
\; , \nonumber \\
v_{\rm cdm}^{^{\rm rad}}(\kappa)
&=&
2 \sqrt{3} \zeta_0 \frac{1-j_0(\kappa)}{\kappa}
\; ,
\end{eqnarray}
where $\kappa \equiv k\eta/\sqrt{3}$, $\zeta_0$ is the value of $\zeta$ in 
the superhorizon limit, $\gamma_{\rm E}$ Euler's constant, Ci the 
cosine integral and $j_n$ are the spherical Bessel functions, all defined as 
in \cite{AbramowitzStegun}. It is easy to check that this is the isentropic 
mode, since in the superhorizon limit $\Delta_{\rm cdm} \to \Delta_{\rm r}$ 
and $v_{\rm cdm} \to v_{\rm r}$.

On superhorizon scales ($k\ll{\cal H}$),  
$\phi \simeq -2\zeta_0/3$, $\Delta_{\rm r} \simeq \zeta_0$,
$v_{\rm r} \simeq (k/{\cal H}) \zeta_0/3$.  
On subhorizon scales ($k\gg{\cal H}$), a first order expansion
of the exact solutions, (\ref{gradex}) and (\ref{gmatex}), in $\kappa$ gives
\begin{eqnarray}
\label{gradap}
\phi^{^{\rm rad}}(\kappa)
&\simeq&
2 \zeta_0 \, \frac{{\rm cos}(\kappa)}{\kappa^2}
\; , \nonumber \\
\Delta_{\rm r}^{^{\rm rad}}(\kappa)
&\simeq&
- 3\zeta_0 \, {\rm cos}(\kappa)\; , \hspace{0.5cm}
\label{lorv}
v_{\rm r}^{^{\rm rad}}(\kappa)
\simeq
\sqrt{3} \, \zeta_0{\rm sin}(\kappa)
\; , \\
\label{gmatap}
\Delta_{\rm cdm}^{^{\rm rad}}(\kappa)
&\simeq&
6\zeta_0\left[{\rm ln}(\kappa) + \gamma_{\rm E} -\frac{1}{2}\right]
\; , \hspace{0.5cm}
v_{\rm cdm}^{^{\rm rad}}(\kappa)
\simeq
\frac{2\sqrt{3}\zeta_0}{\kappa}
\; .
\end{eqnarray}
During radiation domination, the radiation density and velocity
perturbations on subhorizon scales oscillate with constant amplitude,
thereby generating a Newtonian gravitational potential that decays
like $1/\kappa^2$, while the CDM density perturbations grow
logarithmically and the velocity perturbations decay with
$1/\kappa$. From (\ref{gradap}) it can be seen, that the Newtonian
gravitational potential acts like an external field on the evolution
of CDM density perturbations in this regime.

In the next section we discuss the viscous coupling
of WIMPs to radiation and its effect on the evolution of WIMP density
perturbations during radiation domination. While the viscous
coupling has  important consequences for the spectrum of WIMP
perturbations,  the evolution of the radiation fluid is unaffected.

\section{Kinetic decoupling}
\label{kdsec}

Elastic scattering processes around $T_{\rm kd}$ lead to viscous coupling 
of the WIMP and radiation fluids, which results in collisional damping of 
WIMP perturbations, as shown by two of us in \cite{hss}.  
The WIMP perturbations 
disperse due to bulk and shear viscosity.  The strength of these damping 
mechanisms is described in terms of local transport coefficients 
$\zeta_{\rm vis}$ and $\eta_{\rm vis}$ for bulk and shear viscosity.  Close to 
local thermal equilibrium, $\zeta_{\rm vis} \approx 5/3 n T \tau_{\rm relax}$ 
and $\eta_{\rm vis} \approx nT\tau_{\rm relax}$.  The WIMP sound speed in
that regime is given by $\sqrt{(5/3)(T/m)}$. The WIMP perturbations 
that survive the viscous coupling prior to kinetic decoupling are then the 
initial conditions for the free streaming regime.

We now calculate the spectrum of WIMP density perturbations on the surface 
of kinetic decoupling by solving the linearised Navier-Stokes equations. 
For a more general and complete derivation, see \cite{hss}. Dissipation is
a subhorizon phenomenon. The linearised Navier-Stokes equation for the WIMP
fluid on subhorizon scales ($k \gg {\cal H}$), is given by
\begin{equation}
\label{v}
v_{\rm wimp}^{\; \prime} + \frac{\zeta_{\rm vis} + 4/3 \eta_{\rm vis}}
{\epsilon_{\rm wimp}} \; \frac{k^2}{a} v_{\rm wimp} 
= -c_{\rm wimp}^{\; 2} k \Delta_{\rm wimp}, 
\end{equation}
where a gravitational forcing term and energy dissipation due to the
large-scale inhomogeneities of the radiation fluid, as well as horizon scale
contributions, have been dropped. Similarly, the 
continuity equation (\ref{f1}) becomes $\Delta_{\rm wimp}' = k v_{\rm wimp}$. 
These equations can be combined to give a second order differential equation
for the evolution of $\Delta_{\rm wimp}$:
\begin{equation}
\label{osci}
\Delta_{\rm wimp}^{\; \prime\prime} + 
\frac{\zeta_{\rm vis} + 4/3 \eta_{\rm vis}}{\epsilon_{\rm wimp}} 
\; \frac{k^2}{a} \Delta_{\rm wimp}^{\; \prime}
+ c_{\rm wimp}^{\; 2} \; k^2 \; \Delta_{\rm wimp}
= 0.
\end{equation}
Note that the final term should not be neglected as its coefficient
($c^2_{\rm wimp} \sim T/m$) is of comparable magnitude to that of
the dissipative term [$(T/m)(\tau_{\rm relax}/t)$] for $t \sim
\tau_{\rm relax}$.  This is the equation of motion of an oscillator
with time dependent friction, due to the bulk and shear viscosity, and
hence $\Delta_{\rm wimp}$ oscillates with complex frequency. The real
part of the frequency is proportional to the isentropic sound speed in
the WIMP fluid and describes the propagation of perturbations in this
fluid, while the imaginary part describes the damping of the
perturbations caused by the transfer of energy and momentum, due to
the viscous coupling, from the WIMP fluid to the radiation fluid (which
acts as a heat bath). 

Let us stress that our treatment is restricted to $k \gg {\cal H}$. 
Near the horizon there are extra forcing terms on the right hand side 
of (\ref{osci}). There are two types of such terms. Firstly,  
there is the gravitational pull of the oscillating radiation fluid.
Secondly, as long as $c_{\rm wimp} \sim c_{\rm rad}$ the oscillations of
the radiation fluid can drag the CDM fluid along by means of multiple
scatterings in a preferred direction. As soon as the relaxation time
scale exceeds the oscillation period of a given mode, this mechanism can no 
longer be active. This coherent effect is not accounted for by our
viscosity terms, which reflect the dissipation due to the homogeneous 
component of the radiation fluid. There are also some extra terms from
the Hubble expansion on the left hand side of (\ref{osci}). 

The numerical treatment of Loeb and Zaldarriaga \cite{LZ} includes
these terms. Their results confirms that the essential features of the
subhorizon damping are captured by our calculation.  They also show
that the additional terms play an important role in determining the
precise position of the maximum of the CDM power spectrum. We agree
with their conclusion that an accurate calculation (better than $10\%$
accuracy) must be based on a numerical computation including all of the
terms.
 
The WKB solution to (\ref{osci}) for the envelope of the WIMP density 
oscillations is: 
\begin{eqnarray}
\label{solap}
\Delta_{\rm wimp}(k,\eta_{\rm kd})
&=&
\Delta_{\rm wimp}(k,\eta_{\rm i})\;
{\rm exp}\left(-\int_{\eta_{\rm i}}^{\eta_{\rm kd}} 
\frac{\zeta_{\rm vis} + 4/3 \eta_{\rm vis}}{2\epsilon_{\rm wimp}} 
\frac{k^2}{a} {\rm d}\eta \right) \;,
\end{eqnarray}
where $\Delta_{\rm wimp}(k,\eta_{\rm i})$ is the initial primordial
density perturbation and $\eta_{\rm kd}$ is the conformal time at
kinetic decoupling. 
The 
damping term can be written as 
\begin{equation}
\label{dd}
D_{\rm d}(k) \equiv \frac{\Delta_{\rm wimp}(k, \eta_{\rm kd})}
      {\Delta_{\rm wimp}(k, \eta_{\rm i})}
    ={\rm exp} \left[ -\left( \frac{k}{k_{\rm d}} \right)^2 \right] \; , 
\end{equation}
where
$k_{\rm d}$ is given by
\begin{eqnarray}
\label{kd}
k_{\rm d}
&=&
\left(\int_{\eta_{\rm i}}^{\eta_{\rm kd}} 
\frac{\zeta_{\rm vis} + 4/3 \eta_{\rm vis}}{2\epsilon_{\rm wimp} a} 
{\rm d}\eta \right)^{-1/2}
\,, \nonumber \\
&=&
\left(\frac{3}{2}\int_{\eta_{\rm i}}^{\eta_{\rm kd}}
\frac{T}{m}\frac{\tau_{\rm relax}}{a} {\rm d}\eta \right)^{-1/2}
\,, \nonumber \\
&\approx&
1.8 \left(\frac{m}{T_{\rm kd}}\right)^{1/2} \frac{a_{\rm kd}}{a_0} H_{\rm kd}
\,, \nonumber \\
&\approx&
\frac{3.76\times 10^7}{\rm Mpc}\left(\frac{m}{100 \; {\rm GeV}}\right)^{1/2}
\left(\frac{T_{\rm kd}}{30 \; {\rm MeV}}\right)^{1/2}
\; .
\end{eqnarray}
This scale corresponds to a length scale $\sim 10^{-2}/H$ at kinetic
decoupling. The total WIMP mass contained in a sphere with radius
$\pi/k_{\rm d}$ is $M_{\rm d} \sim 10^{-10} M_\odot$. Here we assumed 
a WIMP mass typical for the lightest supersymmetric particle and a 
typical kinetic decoupling temperature for a weakly interacting particle
of such a mass \cite{hss,ghs}. 

Once the WIMPs have kinetically decoupled, they can be described as a
separate fluid. On scales larger than the free streaming length, we
can approximate it as a pressureless fluid. Since kinetic decoupling
happens well before the end of  radiation domination the
gravitational evolution of the WIMP density perturbations on arbitrary
scales during radiation domination ($\epsilon_{\rm m} \ll
\epsilon_{\rm r}$) is given by (\ref{gmatex}).

\section{Free streaming}
\label{fs}

After kinetic decoupling the evolution of the WIMPs at the smallest scales
is described by the distribution function $f({\bf x}, q{\bf n},\eta)$, 
where $q$ and ${\bf n}$ are the norm and direction of the comoving
3-momentum, i.e. $q_{\rm ph} = q/a$. The distribution function  
is governed by the collisonless Boltzmann equation, which reads in a
flat Friedmann model (see e.g. \cite{bern})
\begin{equation}
\label{dfdt}
\left(
\frac{\partial}{\partial\eta}
+
\frac{{\rm d}x^i}{{\rm d}\eta} \frac{\partial}{\partial x^i}
+ \frac{{\rm d}q}{{\rm d}\eta} \frac{\partial}{\partial q}
+ \frac{{\rm d}n_j}{{\rm d}\eta} \frac{\partial}{\partial n_j}
\right) f({\bf x},q{\bf n},\eta)
= 0 \; .
\end{equation} 
In local thermal equilibrium we have $f({\bf x}, q{\bf n},\eta) = f_0(q,\eta)$,
which in the case of non-relativistic particles becomes the Maxwell-Boltzmann 
distribution function (with $g$ internal degrees of freedom), 
\begin{equation}
\label{f0}
f_{\rm MB}(q) = \frac{g}{(2\pi)^3}\; {\rm exp}\left(\frac{\mu-m}{T_{\rm
wimp}}\right)
{\rm exp}\left[-\frac{\left(q/a\right)^2/2m}{T_{\rm wimp}}\right]
\; .
\end{equation}
Here, $T_{\rm wimp}$ denotes the temperature and $\mu$ the chemical potential 
of the WIMPs. Both depend on time, but $f_{\rm MB}(q)$ does not. 

In the next step we have to investigate deviations of the
distribution function of the WIMPs in order to describe an inhomogeneous
universe. We will do so by considering small perturbations in the
Boltzmann equation away from the local thermal equilibrium solution. 
Close to local thermal equilibrium we can write
\begin{equation}
f({\bf x},q{\bf n},\eta)
= f_0(q,\eta) + \delta f({\bf x},q{\bf n},\eta) \; .
\end{equation}   
The last term in (\ref{dfdt}) does not contribute at first order in the 
perturbed quantities. This is because ${\rm d} n_j/{\rm d}\eta$ and
$\partial f/\partial n_j$ are individually first order terms; 
in the absence of metric perturbations free-streaming particles do not 
change direction and the zero-th order distribution function $f_0(q,\eta)$ 
is independent of direction ${\bf n}$. Let us now turn to the third term
in (\ref{dfdt}). {}From the geodesic equation it follows that 
${\rm d}q/{\rm d}\eta$ is proportional to derivatives of metric 
perturbations.  
Free streaming only occurs on scales well below the horizon (due to the small
velocities of WIMPs) and on subhorizon scales the metric perturbations 
$\phi$ and $\psi$ are negligible --- consequently the third term in 
(\ref{dfdt}) can also be neglected.
The collisionless Boltzmann equation for subhorizon scales 
can then be rewritten in terms of the spatial Fourier transform of 
$\delta f({\bf x},q{\bf n},\eta)$, which we denote by 
$\delta f({\bf k},q{\bf n},\eta)$:
\begin{equation}
\label{keqk}
\left(
\frac{\partial}{\partial\eta}
+
i \; \frac{q/a}{m} \; {\bf k}\cdot{\bf n} \right)
\delta f({\bf k},q{\bf n},\eta) 
=
0
\; ,
\end{equation}
Note that the second
term in this equation depends on the direction ${\bf n}$ of the comoving
momentum only via its angle with respect to the wavenumber ${\bf k}$. 

Before solving equation (\ref{keqk}) the initial perturbations have to
be specified. At kinetic decoupling we can assume that $T =
T_{\rm wimp}$ and we drop the index ``wimp'' in the following. 
The WIMP phase space distribution around kinetic decoupling 
is close to a Maxwell-Boltzmann distribution (\ref{f0}).  
The WIMP density perturbations (with ${\cal H}\ll k \ll k_{\rm d}$) 
which are present on the surface of kinetic decoupling need to be taken 
into account.  To do so we consider small fluctuations $\delta T$ and 
$\delta\mu$ of the thermodynamic variables $T$ and $\mu$ in (\ref{f0}) 
and expand the resulting distribution up to first order:
\begin{equation}
\frac{\delta f}{f_0}(k,q,\eta_{\rm kd})
=
\delta \left( \frac{\mu}{T} \right) + \left(\frac{m}{T}
+ \frac{(q/a)^2/2m}{T}\right) \frac{\delta T}{T}
+ {\cal O}\left(\delta^2\right)
\; .
\end{equation}
As we are interested only in effects of first order in the perturbed 
quantities, the adiabatic equations of motion can be used to establish 
the relationships between temperature and chemical potential fluctuations 
as well as WIMP perturbations on the surface of kinetic decoupling. 
{}From the conservation of entropy per WIMP  we find $\delta n/n 
\approx (3/2) \delta T/T$ and thus $\delta T/T \approx 2
\Delta_{\rm wimp}(k,\eta_{\rm kd})/3$.  Inserting this in the Gibbs-Duhem
relation $\delta(\mu/T) = \delta P/(nT) - (\epsilon + P)/(nT)
(\delta T/T)$ and using the previous relation, we find 
$\delta (\mu/T) \approx - (2m/3T) \Delta_{\rm wimp}(k,\eta_{\rm kd})$, so 
that the deviation from local thermal equilibrium at kinetic decoupling
is given by
\begin{equation}
\label{dfin}
\left(\frac{\delta f}{f_0}\right)
(k,q,\eta_{\rm kd})
=
\frac{q_{\rm kd}^2/2m}{3T_{\rm kd}/2}
\Delta_{\rm wimp}(k,\eta_{\rm kd}) 
+ {\cal O}(\delta^2)
\; .
\end{equation}
This has an obvious interpretation, the fractional perturbation of the
distribution function is proportional to the ratio of the kinetic energy
of an individual particle with comoving velocity $q/m$ and the thermal
energy $3T/2$ times the density contrast. 

The solution to the free streaming equation (\ref{keqk}) is then 
\begin{eqnarray}
\label{solfs}
\left(\frac{\delta f}{f_0}\right)({\bf k},q{\bf n},\eta) &=&
\left(\frac{\delta f}{f_0}\right)
(k,q,\eta_{\rm kd})
\; 
{\rm exp}\left[-{\rm i} l(q,\eta) \; {\bf k}\cdot{\bf n}\right]
\; ,
\end{eqnarray}
where $l=l(q,\eta)$ is the comoving distance a WIMP can travel freely 
in the background space-time during the time interval $\eta-\eta_{\rm kd}$:
\begin{equation}
l(q,\eta)
=
\int_{\eta_{\rm kd}}^{\rm \eta} {\rm d}\eta^\prime\; 
\frac{q/m}{a(\eta^\prime)}
\; .
\end{equation} 
We can evaluate that this integral for the matter-radiation universe,
in which we can write the scale factor as
\begin{equation}
a(\eta) = a_{\rm eq} 
\left[1/4 \left(\frac{{\cal H}_{\rm eq}\eta}{\sqrt{2}}\right)^2
+ \frac{{\cal H}_{\rm eq} \eta}{\sqrt{2}}\right] \,.
\end{equation}
Using the previously introduced notation $y = a/a_{\rm eq}$ some algebra
finally yields,
\begin{equation}
\label{l}
l(q,\eta)
= \frac{q_{\rm kd}}{m} y_{\rm kd} 
{\rm ln}\left[\frac{y_{\rm kd}}{y}
              \frac{y+2\left(1-\sqrt{1+y}\right)}
                   {y_{\rm kd} +2\left(1-\sqrt{1+y_{\rm kd}}\right)}\right]
\frac{\sqrt{2}}{{\cal H}_{\rm eq}}\; ,
\end{equation}
where $q_{\rm kd} \equiv q/a_{\rm kd}$ is the modulus of the physical momentum
at kinetic decoupling.  As $y \ll 1$, the physical distance of travel of
free streaming particles 
approaches $[(q_{\rm kd}/m) y_{\rm kd}] \sqrt{2} \ln(4/y_{\rm kd}) 
[a/{\cal H}_{\rm eq}]$, the first square bracket is the physical velocity at 
kinetic decoupling and the scale is set by the physical size of the Hubble 
horizon at equality (the other term in square brackets). 

The free streaming of the WIMPs generates a further (collisionless) 
damping mechanism for WIMP density perturbations.  After averaging over
the Maxwellian distribution of velocities, we can estimate the damping
scale by replacing the physical velocity of an individual WIMP by the
mean velocity (thermal velocity) of WIMPs at kinetic decoupling. Thus
the comoving length scale of free streaming 
\begin{equation}
\label{lfs}
l_{\rm fs} \sim \bar{v}_{\rm kd} a_{\rm kd} \int_{\eta_{\rm kd}}^{\rm \eta}
\frac{{\rm d}\eta^\prime}{a(\eta^\prime)}\; ,
\end{equation} 
with $\bar{v}_{\rm kd} = \sqrt{3 g T/m}$. 
In order to characterise the spectrum of perturbations which survive
collisionless damping we introduce the comoving free streaming scale $k_{\rm
fs}\propto 1/l_{\rm fs}$. More precisely we define 
\begin{equation}
\label{kfs}
k_{\rm fs}(\eta) \equiv {\sqrt{2} \over 
\sqrt{T_{\rm kd}/m} [l(q_{\rm kd},\eta)/(q_{\rm kd}/m)]} \,,  
\end{equation}
where the numerical prefactor is chosen for later convenience. 
With the help of (\ref{l}) we find that the comoving $k_{\rm fs}$ 
becomes approximately constant as $a/a_{\rm eq}\gg 1$ and is given by
\begin{eqnarray}
k_{\rm fs} &\approx & \left({m\over T_{\rm kd}}\right)^{1/2} 
{a_{\rm eq}/a_{\rm kd}\over \ln (4a_{\rm eq}/a_{\rm kd})}\frac{a_{\rm
eq}}{a_0} H_{\rm eq} \,, \nonumber \\
&\approx &
\frac{1.70 \times 10^6}{\rm Mpc}
\frac{(m/100 \; {\rm GeV})^{1/2} (T_{\rm kd}/30 \; {\rm MeV})^{1/2}}
{1+ {\rm ln}(T_{\rm kd}/30 \; {\rm MeV})/19.2}
\; .
\end{eqnarray}
This expressions depends on $\omega_{\rm m} \equiv \Omega_{\rm m} h^2$
only through the logarithm, we therefore set it equal to the WMAP best
fit value, $\omega_{\rm m} = 0.14$ \cite{sperg}. The corresponding
length scale at matter-radiation equality is $\sim 10^{-8}/H$ and the
total WIMP mass contained in a sphere of radius $\pi/k_{\rm fs}$ is
$M_{\rm fs}\sim 10^{-6} M_\odot$.

The WIMP density contrast is related to the distribution function by:
\begin{equation}
\label{delm}
\Delta_{\rm wimp}({\bf k},\eta)
=
\frac{
\int_{0}^{\infty} {\rm d}q q^2 \; 
\int {\rm d}\Omega \; \delta f({\bf k},q{\bf n},\eta)
}{4\pi \int_{0}^{\infty} {\rm d}q q^2 f_0(q,\eta)}
\; ,
\end{equation}
with ${\rm d}q q^2 {\rm d}\Omega$ denoting the volume measure in
spherical coordinates in ${\bf q}$-space.  
The denominator in (\ref{delm}) is the comoving mass density
of CDM and the numerator is the CDM mass fluctuation due to the 
deviation from local thermal equilibrium. Using (\ref{dfin}) and 
(\ref{solfs}) in (\ref{delm}), we find after integration over $\Omega$
\begin{eqnarray}
\label{detail}
\Delta_{\rm wimp}({\bf k},\eta)
&=&
\left(\frac{2}{\pi}\right)^{1/2}
\Delta_{\rm wimp}({\bf k},\eta_{\rm kd})
\left(m T_{\rm kd}\right)^{-3/2}
\int_0^\infty {\rm d}q_{\rm kd} q_{\rm kd}^2 \;
\frac{q_{\rm kd}^2}{3mT_{\rm kd}}
\nonumber \\
&&\times
\exp\left[-\frac{q_{\rm kd}^2}{2mT}\right] \;
\frac{\sin\left(l(q_{\rm kd},\eta) k\right)}
     {l(q_{\rm kd},\eta)k}\; .
\end{eqnarray}
The remaining integration can be found as expression 3.952(5) in 
Reference \cite{GR}.

As a result of the integration we find the suppression of the WIMP density 
contrast due to free streaming to be
\begin{eqnarray}
\label{dfs}
D_{\rm fs}(k,\eta)
&\equiv&
\frac{\Delta_{\rm wimp}({\bf k},\eta)}{\Delta_{\rm wimp}({\bf k},\eta_{\rm kd})}
=
\left[1-\frac{2}{3} \left(\frac{k}{k_{\rm fs}}\right)^2\right]
\;
{\rm exp}\left[-\left(\frac{k}{k_{\rm fs}}\right)^2\right]
\; .
\end{eqnarray}
We see that the free streaming of WIMPs results in exponential
damping of the WIMP density contrast, similar to that produced by collisional
damping (justifying the terminology `collisionless damping' for the
effects of free streaming). There is one difference, however; the
ratio of the WIMP kinetic energy to the thermal averaged
kinetic energy in (\ref{dfin}) leads to a polynomial pre-factor. This
expression is valid for $k/k_{\rm fs} < 1$, as terms of order
$(k/k_{\rm fs})^4$ have been neglected in the polynomial.

The net damping is the product of the collisional
(due to viscous coupling to the radiation fluid) and collisionless
(due to free streaming) damping terms [(\ref{dd}) and (\ref{dfs}) respectively],
$D(k)= D_{\rm d}(k) D_{\rm fs}(k)$:
\begin{equation}
\label{dk}
D(k) 
\equiv 
\frac{\Delta_{\rm wimp} (k, \eta)}{\Delta_{\rm wimp}(k,\eta_{\rm i})}
=
\left[1-\frac{2}{3} \left(\frac{k}{k_{\rm fs}}\right)^2\right]
\;
{\rm exp}\left[-\left(\frac{k}{k_{\rm fs}}\right)^2-
\left(\frac{k}{k_{\rm d}}\right)^2\right]
\; .
\end{equation}
Since $k_{\rm fs} \ll k_{\rm d}$, the cut-off in the power spectrum
is determined by the free streaming scale $k_{\rm fs}$. 

\begin{figure} 
\begin{center} 
\epsfxsize=6.in
\epsfbox{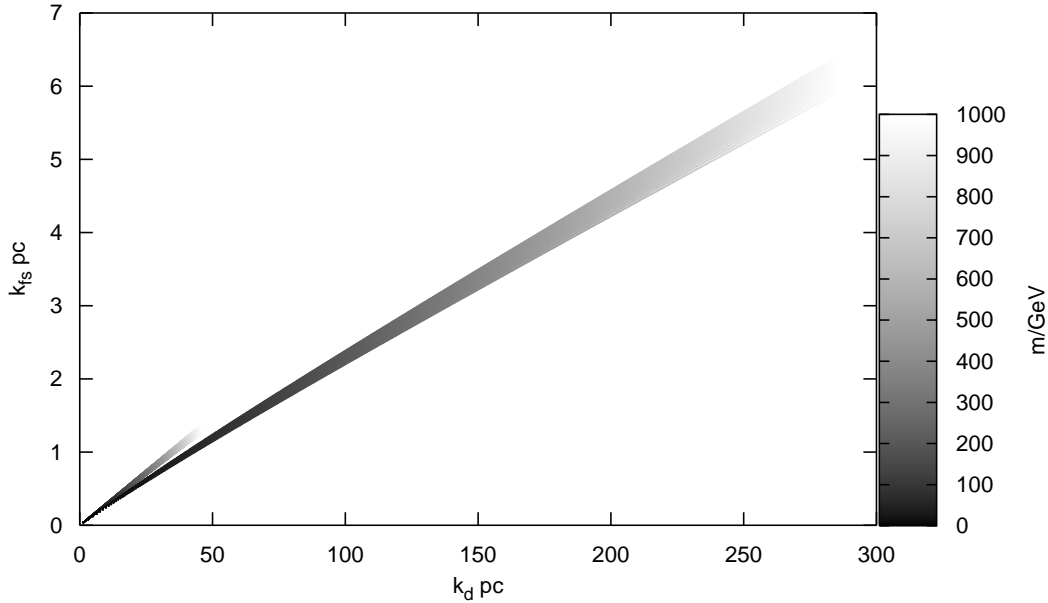} 
\end{center} 
\caption{The variation of the characteristic damping and free
streaming comoving wavenumbers, $k_{\rm d}$ and $k_{fs}$, with WIMP
mass (indicated by grey scale). 
The lower and upper bands are for l=1 (Majorana) and 0 (Dirac) 
respectively.}
\label{kdkfs} 
\end{figure}

In figure~\ref{kdkfs} we plot the variation of the characteristic
damping and free streaming comoving wave numbers with WIMP mass,
for WIMPs with relic
densities corresponding to the WMAP measurement of the CDM
density (i.e.~assuming that the cold dark matter is entirely in the
form of WIMPs) $\omega_{\rm cdm}= \Omega_{\rm cdm}
h^2=0.076 - 0.156$, consistent with the 2-sigma error of WMAP \cite{sperg}. 
For concrete calculations we
will concentrate on four benchmark models which span the range of
most plausible WIMP properties. The details of these benchmark models,
including the values of $k_{\rm d}$ and $k_{\rm fs}$, are tabulated in
Table~\ref{WIMPtable}. Models B and C are very close to the bino-like 
neutralino cases of our previous work \cite{ghs}. Models A and D show
that there is more spread of the predicted damping scale if the strong
assumption of a bino-like WIMP is dropped. 

In order to compare our analytic results with the numerical 
result of reference \cite{LZ}, we express the cut-off scales in terms of
matter mass enclosed in a sphere of radius $R = \pi/k_{\rm fs}$, 
$M_{\rm cut}(R) \approx 4.9 \times 10^{-6} M_\odot (1/k \, {\rm pc})^3$. 
For a decoupling temperature of $25$ MeV, we obtain from \cite{LZ} 
a mass of $6.4 \times 10^{-6} M_\odot$, which is to be contrasted with
$1.5 \times 10^{-6} M_\odot$ from our calculation. So, the difference between
our analytic result and the numerical result including effects near the
horizon scale at decoupling is a factor of a few, instead of two orders of 
magnitude as claimed in \cite{LZ}.

The approximate equality of the kinetic decoupling and free
streaming scales (to within an order of magnitude) can be understood by
comparing the corresponding physical length scales at equality (after
equality the comoving free streaming scale only grows logrithmically).
The free streaming length (\ref{lfs}) is given by the product of
the WIMP velocity at equality, the logarithmic growth factor and the
Hubble radius at equality:  
\begin{equation}
l_{\rm fs} \sim \bar{v}_{\rm
kd} \frac{a_{\rm kd}}{a_{\rm eq}} \ln{\frac{a_{\rm eq}}{a_{\rm kd}}} R_{\rm eq} \,.
\end{equation}
{}From (\ref{kd}), the collisional damping length at
kinetic decoupling is given by the square root of the product of the
viscosity terms with the time of kinetic decoupling, divided by the WIMP
mass density.  The viscosity terms are proportional to the product of the
WIMP mass density, the WIMP sound speed $c_{\rm wimp}$ squared and the
relaxation time.  At kinetic decoupling the relaxation time is by
definition of order the physical Hubble radius. The collisional
damping length at equality is thus 
\begin{equation}
l_{\rm d} \sim c_{\rm wimp} R_{\rm kd} \frac{a_{\rm eq}}{a_{\rm kd}} 
     \sim c_{\rm wimp} \frac{a_{\rm kd}}{a_{\rm eq}}
R_{\rm eq} \,, 
\end{equation}
where we have used $H\propto 1/a^2$ during 
radiation domination and $c_{\rm wimp} \sim \sqrt{T_{\rm kd}/m}$.
At kinetic decoupling the sound speed and the average WIMP velocity
are approximately equal and therefore
$l_{\rm fs} \approx l_{\rm d}  \ln{(a_{\rm eq}/a_{\rm kd})}$, i.e.
the length scales are roughly equal at equality.

This can also be seen from the Reynolds number
of the CDM fluid which is given by Re $\equiv 2 R_H c_{\rm wimp} 
\rho/\eta \sim (R_H/ \tau) c_{\rm wimp} m/T$. At kinetic decoupling we 
have Re $\sim \sqrt{m/T} \sim 1/c_{\rm wimp}$.  Thus the sound speed of
the CDM (equivalent to the mean particle velocity in the
non-relativistic case) at kinetic decoupling plays a
fundamental role in both processes. Kinetic effects dominate over
friction once Re is large. Thus for $T_{\rm kd} \sim $ 10 MeV and 
$m = O(100 {\rm GeV})$ we find Re $\sim 100$. We note that by
fluid engineering standards, this is still a pretty small Reynolds
number. From this consideration we can also see that kinetic decoupling
is a very fast process, as Re $\sim T^{7/2}$ (for the case $l =1$ in Eq.
\ref{xkd}).

\begin{table}[t]
\begin{center}
\caption{Benchmark WIMP models.}
\label{WIMPtable}
\medskip
\begin{tabular}{|c|c|c|c|c|c|c|}
\hline
Ref. & $l$& $m$ (GeV)
& $T_{\rm cd}$ (GeV) & $T_{\rm kd}$ (Mev)& $k_{\rm d} \, ({\rm pc}^{-1})$
&  $k_{\rm fs}  \,({\rm pc}^{-1})$\\
\hline
A &0&  100 & 3.6 & 1.6 & 14 & 0.42 \\
B & 1 & 50 & 1.9 & 21 & 39 & 0.94 \\
C & 1 & 100 & 3.7 & 25 & 61 & 1.5 \\
D & 1 & 500 & 17 & 37 & 180 & 4.0 \\
\hline
\end{tabular}
\end{center}
\end{table}

\section{Evolution of CDM perturbations}
\label{evol}
Now we turn to the gravitational growth of CDM perturbations. So
far we have calculated the damping of the WIMP density perturbations
due to collisional and collisionless damping (\ref{dk}), and the
gravitational evolution of CDM perturbations 
for $\epsilon_{\rm r}\gg\epsilon_{\rm cdm}$,
(\ref{gmatex}). What remains is to find a solution for the
gravitational evolution of CDM perturbations during the matter and
dark energy dominated epochs.  In the following we find analytic
approximations to the radiation-matter and matter-dark energy transitions,
which finally can be smoothly joined with (\ref{gmatex}) to provide
the correct normalisation to CMB measurements.

Since we are interested in the subgalactic scales only, we can
restrict our attention to the subhorizon modes when $\epsilon_{\rm
cdm}$ becomes comparable to $\epsilon_{\rm r}$.  We include neutrinos
in the radiation component in order to allow an analytic
treatment, i.e.\ their anisotropic stress is neglected.  This leads to
errors of around 10 \%~\cite{hu}. We also neglect the baryon
inhomogeneities.  At early times the baryons are tightly coupled to
the radiation fluid, and photon diffusion damping rapidly erases
small-scale perturbations in the baryon fluid at $z \sim 10^6$ to
$10^5$. On small scales the tight coupling breaks down prior to
recombination, and the baryon perturbations grow, however $\Delta_{\rm
b} \ll \Delta_{\rm cdm}$ still~\cite{YSS97,YSS98}. Post decoupling on scales
$k > k_{\rm b} \sim 10^{3} {\rm Mpc}^{-1}$ the residual electrons
allow transfer of energy between the photon and baryon fluids so that
thermal pressure prevents the baryon perturbations from growing, until
$ z_{\rm b} \sim 150$~\cite{YSS97,pad}. 
As we are interested in CDM perturbations on small scales at
early times, we can neglect the perturbations in the baryon fluid.

For scales which enter the horizon sufficiently long before
matter-radiation equality the logarithmic growth of CDM perturbations
provides a situation in which, after some time, $\epsilon_{\rm m}
\Delta_{\rm m} \gg \epsilon_{\rm r} \Delta_{\rm r}$ (i.e. only the CDM
perturbations are important) even during radiation domination
\cite{wein}. We also have $\epsilon_{\rm m} \Delta_{\rm m} \gg
\epsilon_{\rm de} \Delta_{\rm de}$, where the index de stands for dark
energy.  In the following we will assume a cosmological constant, for
which $\Delta_{\rm de} \equiv 0$. Thus we only need to keep
$\Delta_{\rm m}$ in $\Delta$ on the rhs in the Poisson equation
(\ref{e1}).

With these two approximations (subhorizon scales and cdm 
fluctuations dominating 
as the source in the Poisson equation) we can simplify the equations 
(\ref{f1}) and (\ref{f2}) to 
\begin{equation}
\Delta_{\rm cdm}' = k v_{\rm cdm}, \quad v_{\rm cdm}' + {\cal H} v_{\rm cdm} 
= - k \phi \,,
\end{equation}
and the Poisson equation reads
\begin{equation}
-k^2 \phi = 4\pi G a^2 \epsilon_{\rm cdm} \Delta_{\rm cdm} \,.
\end{equation}
It is now most convenient to combine these equations to a single one and to 
use the scale factor instead of conformal time:
\begin{equation}
\label{mpert}
a^2 \frac{{\rm d}^2 \Delta_{\rm cdm}}{{\rm d}a^2} + 
\frac 3 2 \left(1 - \frac P\epsilon\right)
a\frac{{\rm d}\Delta_{\rm cdm}}{{\rm d}a} 
- \frac 3 2 \frac{\epsilon_{\rm cdm}}{\epsilon} \Delta_{\rm cdm} = 0 \,.
\end{equation} 

\subsection{Radiation-matter transition}
\label{sec61}

For the radiation-matter transition (at which point the dark energy,
or cosmological constant, is negligible) (\ref{mpert}) 
simplifies further~\cite{HuS}:
\begin{equation}
\label{meszeq}
y(1+y) \frac{\rm d^2}{{\rm d}y^2} \Delta_{\rm cdm}
+ \left(1+\frac{3}{2} y\right)\frac{\rm d}{{\rm d}y} \Delta_{\rm cdm}
- \frac{3}{2} (1-f_{\rm b}) \Delta_{\rm cdm}
=
0
\; ,
\end{equation}
where $y = a/a_{\rm
eq} = \epsilon_{\rm m}/\epsilon_{\rm r}$ and
$f_{\rm b}=\omega_{\rm b}/ \omega_{\rm m}$ is 
the baryon fraction, with best fit value from WMAP $f_{b}=0.17$~\cite{sperg}. 
The exact solution to this equation is a combination 
of Legendre functions of the first and second kind:
\begin{equation}
\label{meszd}
\Delta_{\rm cdm}(k,y) = B_{1}(k) P_{\nu}(\sqrt{1+y}) + B_{2}(k)
           Q_{\nu}(\sqrt{1+y}) \,,
\end{equation} 
with index $\nu(f_{b})=(\sqrt{25 - 24 f_{b}}-1)/2$. 

We now determine the constants $B_{1,2}(k)$ by matching the $y\ll 1$
expansion of equation (\ref{meszd}) to the
subhorizon limit of the radiation domination solution
(\ref{gmatap}). We find
\begin{eqnarray}
B_1(k)
&=&
6\zeta_0 \left[
{\rm ln}\left(\frac{k}{k_{\rm eq}}\right) + b\right]
\; ,\\
B_2(k)
&=&
- 12\zeta_0
\; ,
\end{eqnarray}
where $k_{\rm eq} \equiv {\cal H}_{\rm eq}$ and
\begin{equation}
b(f_{\rm b}) = \frac{1}{2} \ln\left(\frac{2^5}{3}\right) 
       - \gamma_{\rm E} - \frac{1}{2} - 
    \frac{2}{\nu} - 
 \frac{2\Gamma^\prime[\nu]}{\Gamma[\nu]} \,, 
\end{equation} 
where $\Gamma^\prime[\nu]$ is the derivative of $\Gamma(\nu)$ with respect
to $\nu$.

Finally expanding (\ref{meszd}) for $y\gg1$ we find
that during matter domination,  for scales that enter the horizon
significantly before matter-radiation equality and $z<z_{\rm b}$
\begin{eqnarray}
\label{sol2}
\Delta_{\rm cdm}(y)=  6 \zeta_0\, c(\nu)\, y^{\nu/2}
\left[\ln\left(\frac{k}{k_{\rm eq}}\right) + b\right] \,,
\end{eqnarray}
where
\begin{equation}
c(\nu) = \frac{\Gamma[1+2 \nu]}{2^\nu\Gamma^2[1+\nu]} \,,
\end{equation}
e.g. $\nu(f_{\rm b}) = 1.79\, (2)$, $c[\nu(f_{\rm b})] = 1.36 (3/2)$ and 
$b(f_{\rm b}) = - 1.56 \, (-1.74)$ for $f_{\rm b} = 0.17 \, (0)$.
Note that before $z_{\rm b}$, CDM density perturbations
grow as $\Delta_{\rm cdm} \propto a^{\nu/2}$. Later the baryons follow the 
CDM and the matter fluctuations grow as $a$. For the peculiar velocity and 
the Newtonian gravitational potential we obtain
\begin{eqnarray}
\label{sol2b}
v_{\rm cdm}(y) &=& \frac{k_{\rm eq}}{k} 
\sqrt{\frac{y}{2}} \frac{\rm d}{{\rm d} y}
\Delta_{\rm cdm}(y)
\; , \nonumber \\
\phi(y) &=& -\frac{3}{4} \left(\frac{k_{\rm eq}}{k}\right)^2 (1-f_{\rm b})
{\Delta_{\rm cdm}(y)\over y}.
\end{eqnarray}
In the following, we omit the subscript `cdm'.

For redshifts $z_{\rm eq} > z > z_{\rm b}$ (between matter-radiation 
equality and the epoch at which small-scale baryon perturbations start 
growing) we find the transfer function for the CDM density perturbations
for modes which satisfy $k > k_{\rm b}$  
\begin{equation}
T_\Delta(k,z) = (6 c)^2 
\left[\ln\frac{k}{k_{\rm eq}} + b\right]^2
\left({1+z_{\rm eq}\over 1+z}\right)^\nu \,,
\end{equation}
and the transfer function for the Newtonian gravitational potential
on these scales is given by
\begin{equation}
T_\phi(k,z) = \left[\frac{27 (1-f_{\rm b}) c}{4}\right]^2
\left[\ln\frac{k}{k_{\rm eq}} + b\right]^2
\left({k_{\rm eq}\over k}\right)^4 
\left({1+z_{\rm eq}\over 1+z}\right)^{\nu-2} \,.
\end{equation}
The transfer function for the velocity depends on the initial time and
is therefore not a very useful quantity. 

\subsection{Normalisation including anisotropic stress from neutrinos
on superhorizon scales} 

Above we have made the assumption that $\phi = \psi$ as we neglect the
anisotropic stress of neutrinos, i.e.~$\pi_{\nu} = 0$. For the
normalisation this leads to unacceptably large errors. We therefore
take the superhorizon anisotropic stress of neutrinos into account in
the normalisation.

Let us define the radiation fraction in neutrinos 
\begin{equation}
\alpha \equiv {\epsilon_{\nu}\over \epsilon_{\rm r}} = 0.405 \,,
\end{equation} 
after $e^+ e^-$ annihilation and for three massless neutrinos. The
regular solution for neutrinos at superhorizon scales during the
radiation dominant epoch gives \cite{neutrinos,RS}
\begin{equation}
\psi_0 = \left(1  + \frac{2\alpha}{5}\right)\phi_0, \quad 
\Delta_{{\rm r}0} = \Delta_{{\rm cdm}0} = - \frac{3}{2} \phi_0 \,, 
\end{equation} 
and thus
\begin{equation}
\zeta_0 = - \frac{3}{2}\left(1 + \frac{4\alpha}{15}\right) \phi_0 \,. 
\end{equation} 
The conservation of $\zeta$ on superhorizon scales then implies  
that, for $z < z_{\rm eq}$ but before the start of dark energy
domination, the normalisation of CDM fluctuations on superhorizon scales
is given as (now we can again assume $\phi = \psi$ since the neutrino 
anisotropic stress is now suppressed by $\epsilon_{\nu}/\epsilon_{\rm m}$)
\begin{equation}
\label{neutrinonormalisation}
\Delta_{\rm cdm} = - \frac{9}{5}\left(1 + \frac{4\alpha}{15}\right)
\phi_0 \,. 
\end{equation} 
When normalising to $\zeta_0$, which is provided by the WMAP measurement
(encoded as the variable $A$), we do not have to take special care as
the WMAP value includes the effect of neutrinos on superhorizon scales.
However, when comparing our analytic result with the results of the
COSMICS code below, we have to include this correction, as the initial
condition in the code is $\phi_0 = -1$. 

\subsection{Accuracy of approximations} 
\label{sec63}

\begin{figure} 
\begin{center} 
\epsfxsize=6.in
\epsfbox{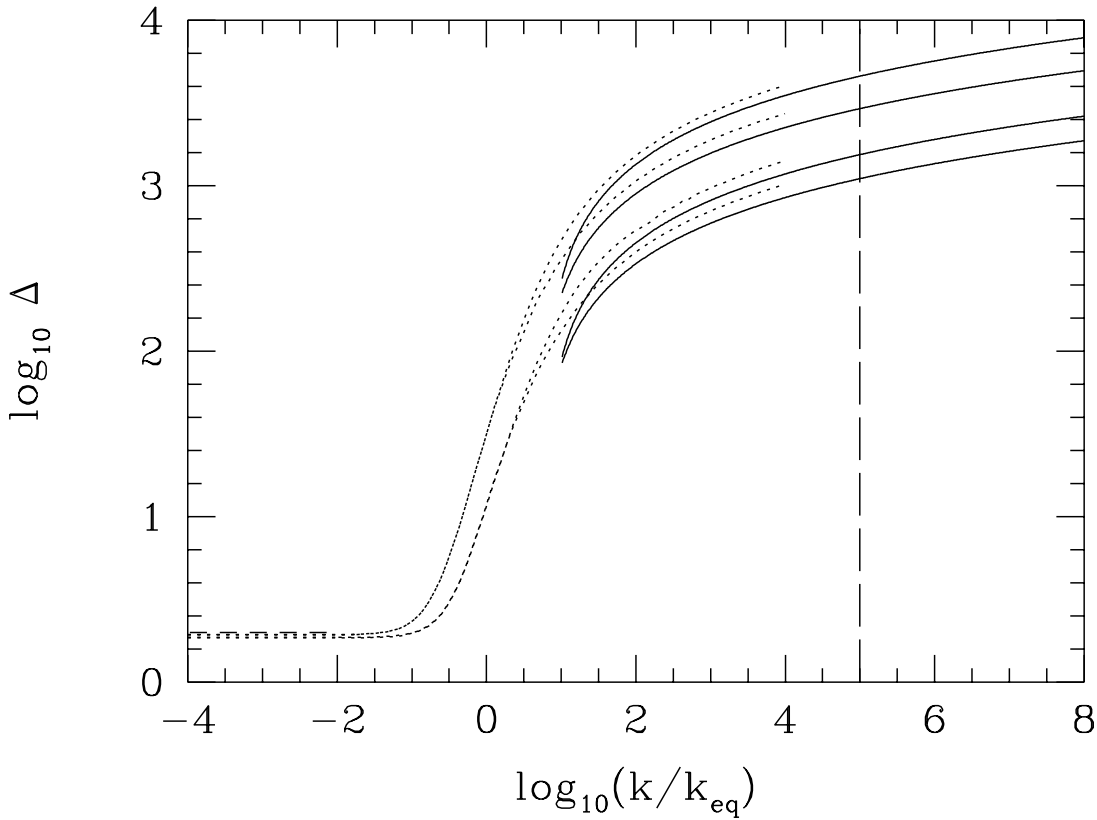} 
\end{center} 
\caption{The CDM density contrast
$\Delta$ at $z=300$ (lower lines) and $z=100$ (upper lines) 
for $\omega_{\rm cdm}=0.116$, $\omega_{\rm b}=0.024$ and 
$\omega_{\rm cdm}=0.14$, $\omega_{\rm b}=0.00$ 
(bottom and top line of each pair respectively), 
using our analytic expression (\ref{sol2}) (solid line) 
and also the COSMICS package (dotted line). 
The analytic superhorizon normalisation
(\ref{neutrinonormalisation}) is shown as a dashed line for 
$k< 10^{-2} k_{\rm eq}$. The
vertical dashed line denotes $k_b$. Our approximation applies best at 
$k>k_{\rm b}$ but, as the comparison shows, it is also pretty good for
smaller wavenumbers.} 
\label{delfig} 
\end{figure}

In Figure~\ref{delfig} we plot the CDM density contrast at $z=300$ and
$100$ as a function of comoving wavenumber for the WMAP best fit total
matter and baryon densities, $\omega_{\rm cdm}=0.116$, $\omega_{\rm
b}=0.024$ ($f_{\rm b}=0.171$ and $b(0.171)=-1.562$), and for zero
baryon density, $\omega_{\rm cdm}=0.14$, $\omega_{\rm b}=0.00$, with
$h = 0.72$, using our analytic expressions and also using the
``lingercon'' Boltzmann solver from the COSMICS package (which
includes massless neutrinos)~\cite{cosmics}. To allow direct
comparison with the output of COSMICS we take initial conditions
(during radiation domination) here such that $\Delta_{\rm cdm}= 3/2$
(i.e. $\phi_0 = -1$).  The rapid oscillations of $\Delta_{\rm r}$ on
sub-horizon scales means that it is not possible to accurately
numerically evolve the coupled perturbation equations for the
sub-galactic scales ($k \gg 10^{4} k_{\rm eq}$) that we are interested
in.  We see that our analytic expression accurately reproduces the
shape of $\Delta$ for $k > 10^{2} k_{\rm eq}$ and the normalisation is
accurate to $ \sim 10\%$.  The accuracy initially increases as $z$
becomes smaller, which can be easily understood from the fact that we
assumed that $z\ll z_{\rm eq}$. Our calculation then becomes less
accurate for $z < z_{\rm b}$.

We also see that baryons have a significant effect on the growth of the 
density contrast; at $z=300$ $\Delta$ is roughly $40\%$ smaller in a 
Universe with the WMAP best fit energy densities ($\omega_{\rm cdm}=0.116$,
$\omega_{\rm b}=0.024$) than in a Universe with the same total matter
density, but no baryons ($\omega_{\rm cdm}=0.14$, $\omega_{\rm
b}=0.00$). 

We conclude that our analytical accuracy is good enough in the light of
the present uncertainties in the cosmological parameters and primordial
power spectrum.

\subsection{Matter-dark energy transition}
\label{secLambda}

Let us stress before going into any details here, that for the
subgalactic scales we are interested in, the following calculation is
not directly relevant, since those modes become non-linear long before
the onset of dark energy domination. However, we need to take the
suppression of perturbation growth into account on larger scales when
we make contact with the measurement of $\sigma_8$ below (section
\ref{results}).

In $\Lambda$CDM cosmologies at late times the cosmological constant,
or dark energy, dominates the Universe, leading to accelerated expansion 
and the suppression of the growth of density perturbations. For this situation 
equation (\ref{mpert}) can be simplified neglecting the presence of radiation. 
We define the scale factor relative to that at the epoch at which the 
matter and cosmological constant densities are equal: $u = a/a_{{\rm eq}2}$ 
and $a_{{\rm eq}2} = a_0 \left(\Omega_{\rm m}/\Omega_{\Lambda}\right)^{1/3}$
and thus $u(a_0) = \left(\Omega_{\Lambda}/\Omega_{\rm m}\right)^{1/3}$.
The evolution of the matter density contrast is now governed by
\begin{equation}
u^2 \frac{\rm d^2}{{\rm d}u^2} \Delta_{\rm m}
+ \frac{3}{2}\left({1 + 2 u^3\over 1 + u^3}\right) 
u\frac{\rm d}{{\rm d}u} \Delta_{\rm m}
- \frac{3}{2} \left({1\over 1 + u^3}\right) \Delta_{\rm m}
= 0 \,,  
\end{equation}
which can be obtained from (\ref{mpert}) using the change of variables 
$\Delta_{\rm m} = u w$ and $u^3 = - z$. The function $w(z)$ obeys a 
degenerate hypergeometric differential equation. Its general solution 
is [see \cite{Erdelyi}, sec. 2.2.2 case 2, eqs. 2.9(1) and 2.9(18) with 
$a=1,b=1/3,c=11/6$]
\begin{equation}
\Delta_{\rm m} = C_1(k)\, u\, 
{}_2F_1(1,{\textstyle\frac 13};{\textstyle \frac{11}6}; -u^3)   
+ C_2(k) \sqrt{1 + \frac{1}{u^3}} \,. 
\end{equation}
The first solution is regular for small $u$, the second one is singular.
We have to join the regular solution to the growing mode 
(\ref{sol2}). In the limit $u \to 0$ we have ${}_2F_1(1,1/3;11/6; -u^3) \to 1$ 
and thus 
$C_1(k) = (3/2)B_1(k)(1+z_{\rm eq})(\Omega_\Lambda/\Omega_{\rm m})^{1/3}$ and
$C_2(k) = 0$. 

We can now put everything together to obtain the 
linear matter density contrast today 
\begin{equation}
\Delta_{\rm m}(k) = 
9\zeta_0(1+z_{\rm eq})
\left(\frac{\Omega_\Lambda}{\Omega_{\rm m}}\right)^{\frac 23} 
\left[\ln\left( \frac{k}{k_{\rm eq}} \right) + b\right] 
{}_2 F_1(1,{\textstyle\frac 13};{\textstyle\frac{11}{6}}; 
- {\textstyle{\Omega_\Lambda\over\Omega_{\rm m}}}) \,. 
\end{equation} 
An excellent analytic approximation to evaluate the hypergeometric 
function for $\Omega_\Lambda/\Omega_{\rm m} > 1$ is given by the 
asymptotic expansion (see \cite{AbramowitzStegun})
\begin{eqnarray}
u\, {}_2 F_1(1,{\textstyle\frac{1}{3}};{\textstyle\frac{11}{6}}; -
u^3) &\to&
{2\Gamma({\textstyle\frac{2}{3}})\Gamma({\textstyle\frac{11}{6}})\over
\sqrt{\pi}} \left(1 + \frac{1}{2 u^3}\right) - \frac{5}{4 u^2} + {\cal
O}(\frac{1}{u^5}) \nonumber \\ &\approx& 1.437 - \frac{1.25}{u^2} +
\frac{0.719}{u^3} + {\cal O}(\frac{1}{u^5}) \,,
\end{eqnarray}
which has an accuracy better than $1\%$ for $\Omega_{\rm m} < 0.4$.  

The growth of structure comes to an end asymptotically; in a pure CDM
model the density contrast grows as $u$, however in a $\Lambda$CDM
universe it only grows by a factor $1.437$ after matter-dark energy
equality.  An equivalent solution has been obtained by Eisenstein
\cite{Eisenstein} in terms of elliptic functions, however he expanded
the elliptical functions for small $u$, which would be more
appropriate for $\Omega_\Lambda/\Omega_{\rm m} < 1$.  One could also
write the above hypergeometric function (or elliptical functions of
Eisenstein) in terms of associated Legendre functions, but it seems to
us that nothing is gained by doing so.

\section{Linear power spectrum}
\label{ps}

In this section we present the linear dimensionless power spectra (defined as
${\cal P}_{X}(k,z) = (k^3/2\pi^2)\langle|X(k,z)|^2\rangle$) normalized
to the WMAP measurements (\cite{sperg,verde}). 

\subsection{Scale invariant spectrum}

For simplicity we start with a scale-invariant primordial power spectrum
and assume that gravitational waves have a negligible contribution to the
CMB anisotropies. We find for $k > k_{\rm b}$ and 
$z_{\rm eq} \gg z \gg z_{\rm b}$
\begin{eqnarray}
{{\cal P}_{\Delta}(k,z)\over 10^{-7} A} 
&=& 
1.06\, c^2 \left[\ln\frac{k}{k_{\rm eq}} + b\right]^2 D^2(k)
\left(1+z_{\rm eq}\over 1+z\right)^\nu \,, \\
{{\cal P}_v(k,z)\over 10^{-7} A} 
&=& 
0.13\, c^2 \nu^2 
\left[\ln\frac{k}{k_{\rm eq}} + b\right]^2 
\left(\frac{k_{\rm eq}}{k}\right)^2 D^2(k)
\left(\frac{1+z_{\rm eq}}{1+z}\right)^{\nu-1} \,, \\
{{\cal P}_\phi (k,z)\over 10^{-7} A} 
&=&  
0.60\, c^2 (1-f_{\rm b})^2 
\left[\ln\frac{k}{k_{\rm eq}} + b\right]^2 
\left(\frac{k_{\rm eq}}{k}\right)^4 D^2(k)
\left(\frac{1+z_{\rm eq}}{1+z}\right)^{\nu-2} \,,
\end{eqnarray}
where $A = 0.9 \pm 0.1$ at the pivot scale $k_0 = 0.05/$Mpc according to 
Reference \cite{sperg}. Note that from the WMAP data the spectral index 
$n = 0.99 \pm 0.04$ is consistent with the scale-invariant 
Harrison-Zel'dovich spectrum $n=1$.  The scale of equality is $k_{\rm eq} =
(0.01/{\rm Mpc}) (\omega_{\rm m}/0.14)$ and $1 + z_{\rm eq} = 3371
(\omega_{\rm m}/0.14)$. 

\begin{figure} 
\begin{center} 
\epsfxsize=6.in 
\epsfbox{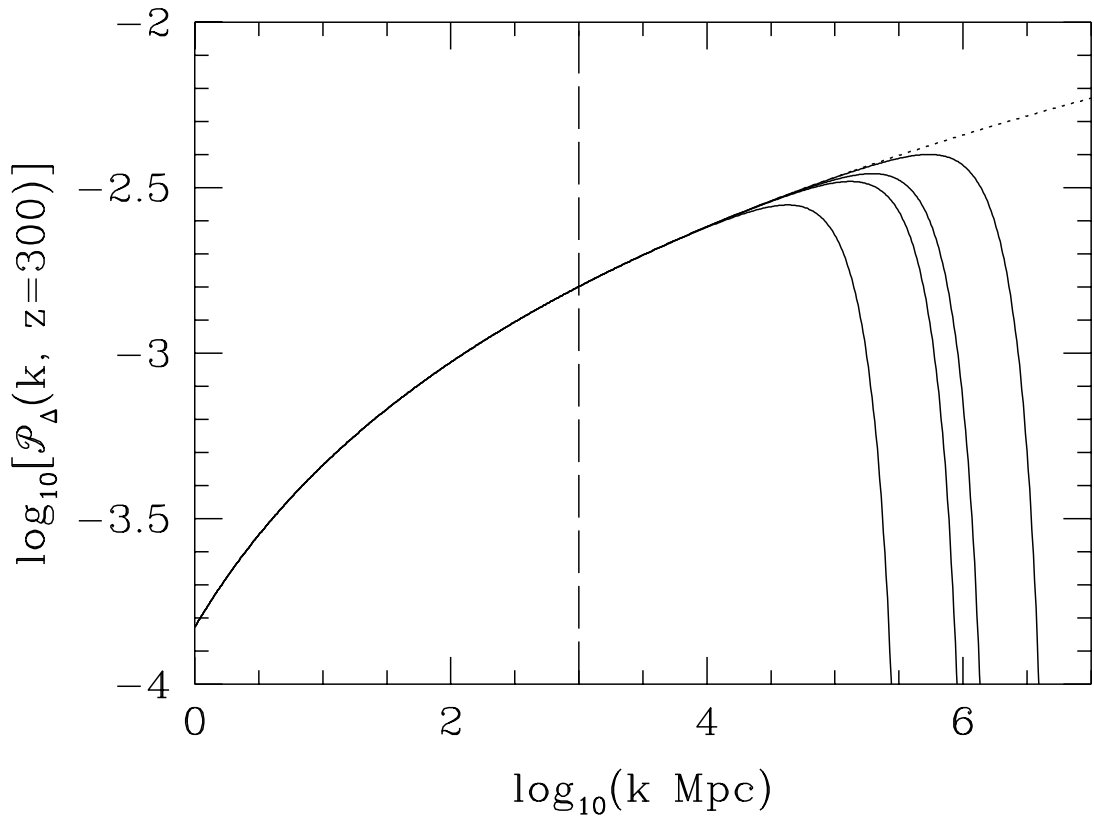}
\end{center} 
\caption{The dimensionless
power spectrum of the WIMP density contrast at $z = 300$ for our four
benchmark WIMP models assuming a scale-invariant primordial power
spectrum (full lines, from left to
right models A, B, C and D). 
Without the effects of collisional damping and free streaming,
the power spectra would be given by the dotted line.
The vertical dashed line denotes $k_b$, the wavenumber below which baryons 
follow CDM. Our approximations are optimised for $k > k_{\rm b}$.  
\label{psdamp}}
\end{figure}

Figure~\ref{psdamp} shows the power spectrum for the WIMP density
contrast at a redshift of $300$, close to the end of the linear regime
of structure formation, with and without the effects of
collisional damping and free-streaming for the four benchmark WIMP models
introduced in Section\ref{fs}.  
In accordance with the WMAP best fit values
we take $\omega_{\rm cdm} = 0.116$ and $f_{\rm b} = 0.17$, and we
assume a scale invariant primordial power spectrum (i.e. $n=1$).
It can be observed that the cut-off of the power spectrum is indeed
very sharp (with a maximum close to the cut-off). As discussed in
section~\ref{fs} the characteristic free-streaming wave-number is
smallest for model A (Dirac WIMP with l=0) and increases with
increasing mass for the Majorana WIMPs (models B-D). This is clearly
reflected in the position of the cut-off in the power spectrum.

\subsection{Scale dependent spectrum} 
\label{pps}

On the scales probed by the CMB [${\cal O} (0.01-0.1) {\rm Mpc}^{-1}$]
the primordial power spectrum is close to scale
invariant~\cite{sperg}. The free-streaming scale $k_{\rm fs} \sim
10^{6} {\rm Mpc}^{-1}$ is seven orders of magnitude smaller and hence
even a very small scale dependence of the power spectrum could
significantly change the amplitude of the power spectrum close to the
cut-off, and hence the red-shift at which the first WIMP halos form.

To assess the effects of possible scale dependence of the primordial
power spectrum we consider three benchmark inflation models which span
the range of possible power spectra for simple inflation models: a
$V=m^{2}\phi^{2}$ chaotic inflation model, power law inflation and a
hybrid inflation model. These models span an interesting region of
inflationary parameter space, for a more detailed discussion see 
\cite{ts}. 

There are also uncertainties in the parameterization of the power
spectrum. The most commonly used parameterization is
\begin{equation}
\label{prim1}
{\cal P}(k)={\cal P}(k_{0}) \left(
          \frac{k}{k_{0}} \right)^{n(k_{0})-1 + \frac{1}{2} 
     \alpha(k_{0}) {\rm ln}(k/k_{0})} \,,
\end{equation}
where $\alpha(k)={\rm d} n/ {\rm d} k$. An arguably more appropriate 
parameterization over a wide-range of scales is~\cite{hf,llms}
\begin{equation}
\label{prim2}
\frac{{\cal P}(k)}{{\cal P}(k_{0})} =
         a_{0} + a_{1} \ln{ \left( \frac{k}{k_{0}} \right)} 
         + \frac{a_{2}}{2}  \ln^2{ \left( \frac{k}{k_{0}} \right)} \,.
\end{equation} 
A small difference between the two parameterizations can be used as an
indicator, that the slow-roll approximation is justified for the model
at hand \cite{llms}. 

The spectral index, $n$, its running $\alpha$ and the alternative
expansion co-efficients $a_{n}$ depend on the inflationary potential
and are most conveniently expressed in terms of the horizon flow
parameters~\cite{hf} which are defined as: 
$\epsilon_{0} \equiv H(N_{i})/ H(N)$, where $N
\equiv \ln({a/a_{i}})$ is the number of e-foldings of inflation since
some initial time $t_{i}$, and 
\begin{equation}
\epsilon_{n+1} \equiv \frac{{\rm d} \ln{ |\epsilon_{n}|}} 
         {{\rm d} N} \,  \hspace{1.0cm} n \geq 1 \,.
\end{equation}
The horizon flow parameters are related to the traditional slow roll
parameters, $\epsilon = (m_{\rm pl}^{2}/ 16 \pi) (V^{'} / 
V)^2$, $\eta = (m_{\rm pl}^{2}/ 8 \pi) (V^{''}/V)$ and $\xi^{2} 
= (m_{\rm pl}^{2}/ 8 \pi)^2 ( V^{'} V^{'''}/ V^2)$ where ${}^{'} 
\equiv {\rm d} / {\rm d} \phi$, as $\epsilon_{1} = \epsilon$,
$\epsilon_{2}= 2 \epsilon - 2 \eta$ and $\epsilon_{2} \epsilon_{3} = 4
\epsilon^{2} - 6 \epsilon \eta + 2 \xi^{2}$.

To first order $n-1=-2 \epsilon_{1} - \epsilon_{2}$ (see
e.g. Ref.~\cite{sg} for the higher order terms) and $\alpha  =  - 2
\epsilon_{1} \epsilon_{2} - \epsilon_{2} \epsilon_{3}$~\cite{hf}.
$a_{0}, a_{1}$ and $a_{2}$ are given by equations (26) -- (28) of 
Ref.~\cite{llms}.

\subsubsection{$m^2 \phi^2$ chaotic inflation}

In this model $\epsilon_{1} = \epsilon_{2}/2 = \epsilon_{3}/2 = 1/(2N+1)$ 
(see e.g. Ref.~\cite{ll}) and we take $N=55$~\cite{Nefolds}. This
gives, including the higher order terms in the expression for $n$,
$n-1=-0.03611$ and $\alpha=-6.49 \times 10^{-4}$.

\subsubsection{Power law inflation}

In power law inflation the scale factor grows as $a \propto t^{p}$
(with $p>1$) and $\epsilon_{1}= 1/p$ with all other horizon flow
parameters equal to zero. We pick $p=55.4$ so that the spectral index
is the same as for the $m^{2} \phi^{2}$ chaotic inflation model
($n-1=-0.03611$).  In this case there is no running of the power
spectrum ($\alpha=0$), i.e.\ the spectral index is constant.

\subsubsection{Hybrid inflation}

Our benchmark hybrid inflation model has potential
\begin{equation}
V=V_{0} \left[ 1 + \frac{1}{2} \left( \frac{\phi}{m_{\rm Pl}} \right)^{2}
     \right] \,,
\end{equation}
and we assume that the first, false vacuum term, in the potential
dominates so that $\epsilon_{1} \ll \epsilon_{2}$, with $\epsilon_{2}$
constant. We take $\epsilon_{2}=-0.014$, the
2-$\sigma$ lower limit from WMAP and 2dF found in Ref.~\cite{ll} giving 
the largest increase in the density contrast with increasing $k$, so
that $n-1=0.03550$ and $\alpha=0$.

\begin{figure}
\begin{center}
\epsfxsize=6.in
\epsfbox{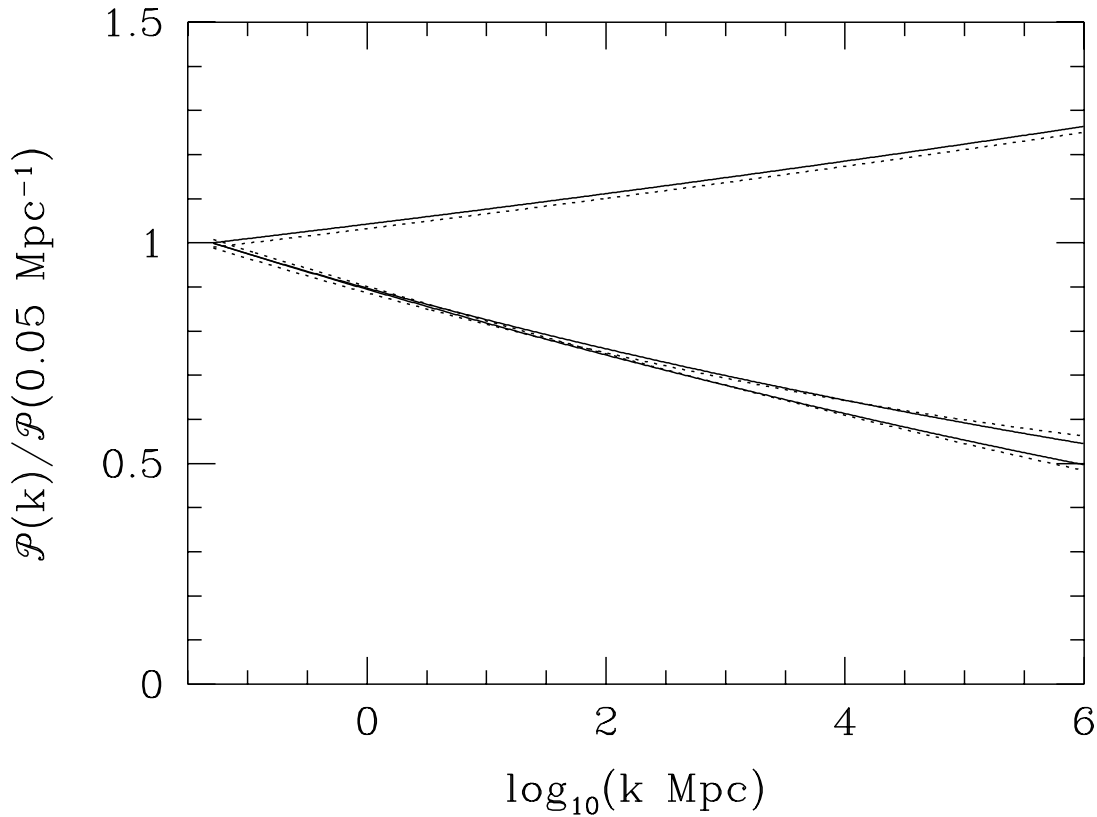} 
\end{center} 
\caption{The primordial power spectra of the three benchmark inflation
models discussed in the text (from top to bottom: hybrid,
power law and $m^2 \phi^2$ chaotic inflation) for the standard
power law parameterisation of the power spectrum (equation (\ref{prim1}), 
solid line) and for the expansion in $\ln{(k/k_{0})}$ (equation (\ref{prim2}),
dotted line).}
\label{primordial}
\end{figure}

The primordial power spectra of these three models are plotted in
Figure~\ref{primordial} for both parameterizations of the power
spectra (equations (\ref{prim1}) and (\ref{prim2})). We see that the
amplitude of the primordial power spectra on the free-streaming scale
$\sim 10^{6} {\rm Mpc}^{-1}$ varies by a factor of $\sim 2.5$
(equivalently the amplitude of the fluctuations varies by $\sim
\sqrt{2.5} \sim 1.6$). We also observe that the two parameterisations
do not give rise to a significant difference, thus we will stick to
the more traditional power-law shape in the following.

\begin{figure}
\begin{center}
\epsfxsize=6.in
\epsfbox{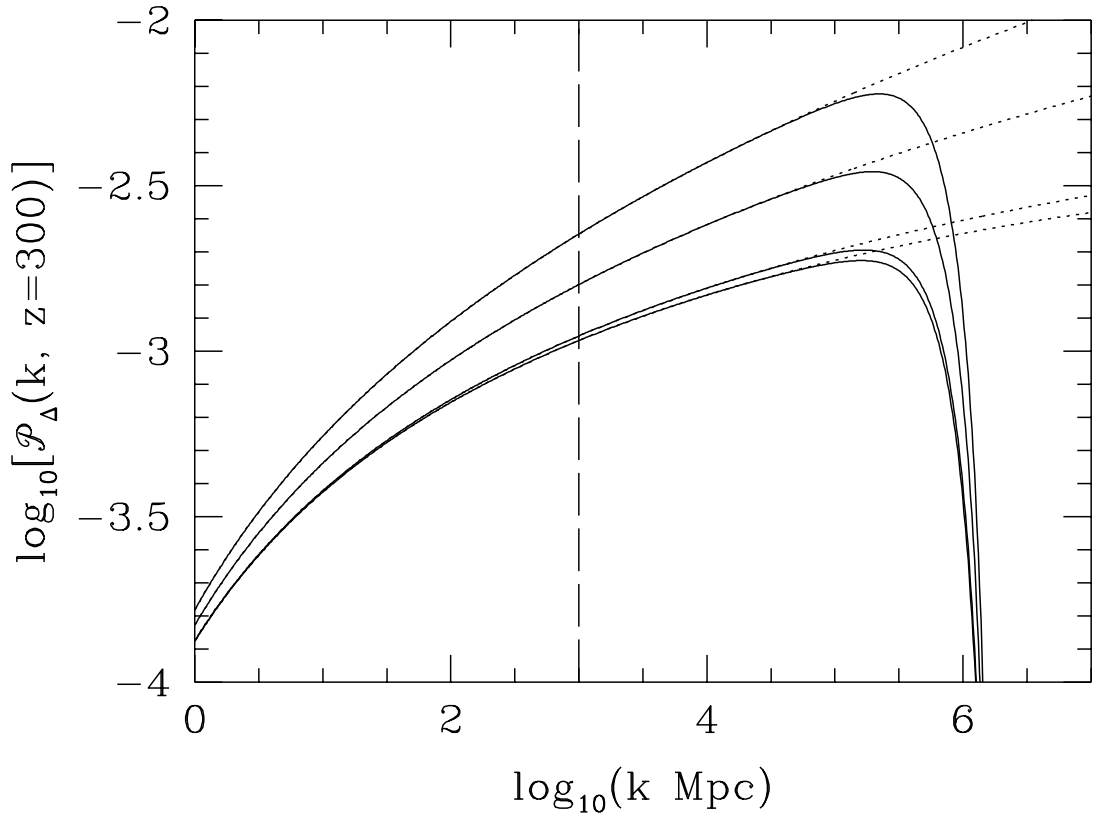}
\end{center}
\caption{The processed power spectra for the three benchmark inflation
models and also a scale invariant primordial power spectrum for 
WIMP benchmark C (from top to bottom: hybrid inflation, scale invariant
primordial spectrum, power law and $m^2 \phi^2$ chaotic inflation).
As before the dotted lines are the power spectra without
the effects of collisional damping and free streaming and
the vertical dashed line denotes $k_b$.} 
\label{psinf}
\end{figure}

In figure~\ref{psinf} we plot the processed power spectra for these
primordial power spectra and also a scale invariant primordial power
spectrum for WIMP benchmark C, with cosmological parameters fixed to
the WMAP best fit values as before. The variation
in the amplitude of the primordial power spectra for $k \sim 10^{6}
{\rm Mpc}^{-1}$ translates directly into a 
variation in the peak amplitude of the processed power spectra.

\section{The first structures}
\label{results}

The collisional damping and free streaming of WIMPs lead
to a cut-off in the WIMP power spectrum, which sets the typical scale
for the first halos in the hierarchical picture of structure
formation. We estimate the redshift at which typical fluctuations on
comoving scale $R$ go nonlinear via
\begin{equation}
\label{znl}
\sigma(R, z_{\rm nl}) = 1, 
\end{equation}  
where $\sigma(R,z)$ is the mass variance defined by
\begin{equation}
\sigma^2(R, z) = \int_{0}^{\infty} W^2(kR) {\cal P}_{\Delta}(k,z)
        \frac{{\rm d} k}{k} \,,
\end{equation}
where $W(kR)$ is the Fourier transform of the window function divided
by its volume. In accordance with the usual procedure, we take the
window function to be a top hat. We normalize $\sigma(R,z)$ to
$\sigma_8 \equiv \sigma(8/h\, {\rm Mpc},0) = 0.9 \pm 0.1$~\cite{sperg}, 
taking into account the suppression of the growth of
$\Delta$ at late times due to the cosmological constant (see section
\ref{secLambda}), as our analytic calculation of the transfer function 
breaks down for modes close to $k_{\rm eq}$. For the purpose of
estimating $z_{\rm nl}$ we ignore the effects of baryons (see section
\ref{sec61} and \ref{sec63}) and assume a $\Lambda$ plus matter universe 
with $\Omega_{\rm m}$ and $\Omega_{\Lambda}$ as 
determined by WMAP~\cite{sperg}. Specifically, we take 
$\Omega_{\rm m}/\Omega_{\Lambda} = 0.370$.  

\begin{figure}
\begin{center}
\epsfxsize=6.in
\epsfbox{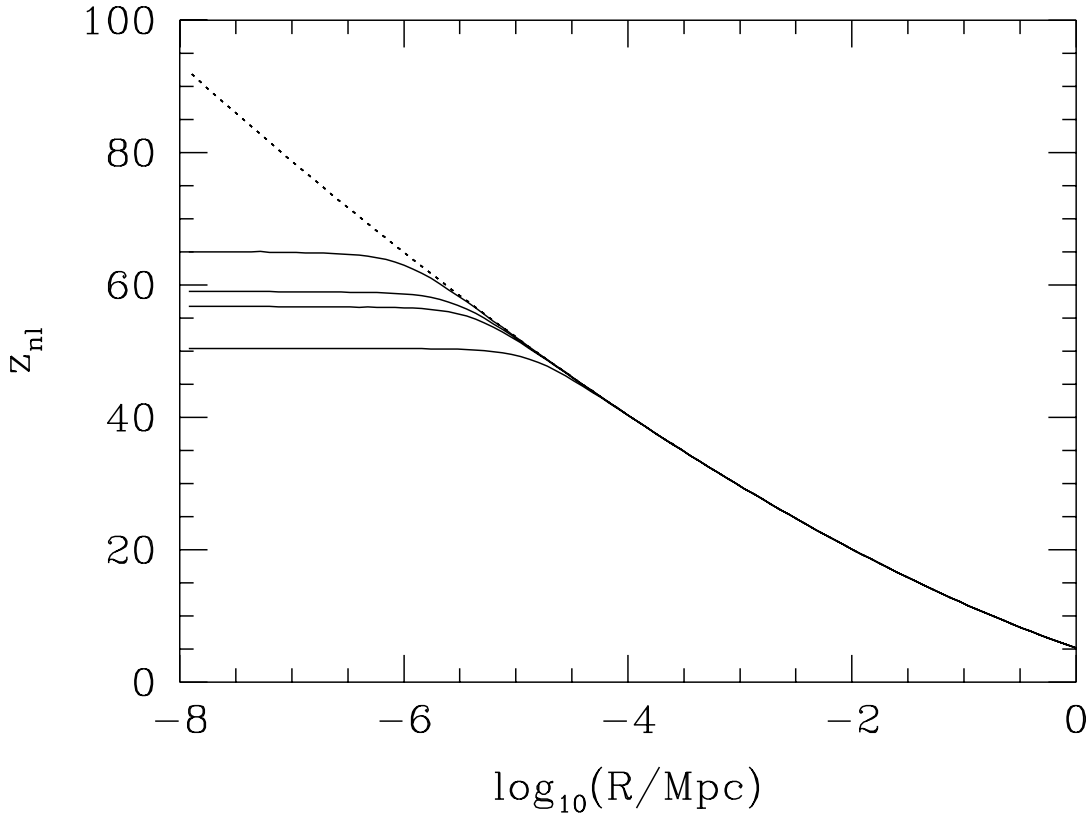}
\end{center}
\caption{The redshift at which typical fluctuations of comoving scale
$R$ become non-linear, $z_{\rm nl}$, 
for the WIMP benchmark models discussed in the text
(from top to bottom: D, C, B, A). The
full lines take into account the effects of collisional damping and
free streaming, whereas the dashed line shows the
behaviour without a cut-off in the power spectrum. The normalisation is
fixed by $\sigma_8 = 0.9$. \label{znldamp}}
\end{figure}

\begin{figure}
\begin{center}
\epsfxsize=6.in
\epsfbox{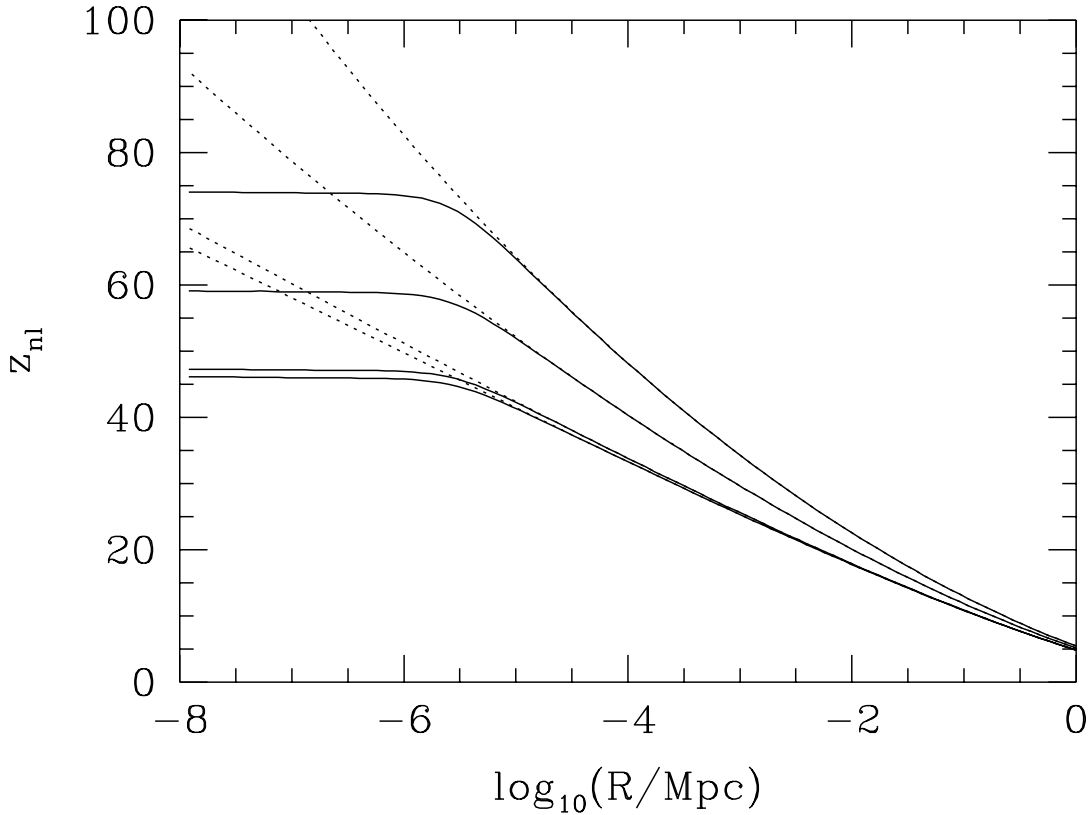}
\end{center}
\caption{The redshift at which typical fluctuations of comoving scale
$R$ become non-linear, $z_{\rm nl}$, now for the primordial power spectra 
discussed in the text (from top to bottom:  hybrid inflation, 
scale invariant primordial spectrum, power law and $m^2 \phi^2$ chaotic 
inflation). The normalisation is fixed by $\sigma_8 = 0.9$. \label{znldamp2}}
\end{figure}

Figures~\ref{znldamp} and \ref{znldamp2} show $z_{\rm nl}$, as defined
by equation~(\ref{znldamp}), as a function of the scale $R$, for the
benchmark WIMPs and primordial power spectra respectively.  In each
case the cut-off in the processed power spectrum at $k \sim 10^{6}
{\rm Mpc}^{-1}$ leads to a plateau with $z_{\rm nl}= z_{\rm nl}^{\rm
max}$ at $R < R_{\rm min} = {\cal O}(1)$ pc. 

We can now give a more precise picture of the onset of the
hierarchical structure formation process; non-linear structure
formation starts at a redshift $z_{\rm nl}^{\max}$. For the
benchmark WIMP models we see that the order of magnitude variation in
$k_{\rm fs}$ leads to a similar variation in $R_{\rm min}$ and also
(because of the dependence of the amplitude of the peak of the power
spectrum on the cut-off scale, see figure~\ref{psdamp}) a range of
values $z_{\rm nl}^{\rm max} \approx 50$ to $65$. For the scale-dependent 
primordial power spectra the factor of $1.6$ (see section \ref{pps}) 
variation in the amplitude of the fluctuations on scales $k \sim 10^{6} 
{\rm Mpc}^{-1}$ translates directly into a comparable range of
values for $z_{\rm nl}^{\rm max}$. Thus, for plausible WIMP
properties and for a range of inflation models which produce
scale-dependent power spectra consistent with the WMAP data 
$z_{\rm nl}^{\max}$ for typical fluctuations takes values in the 
range 40 to 80. For the best fit WMAP matter density and a
scale invariant power spectrum $z_{\rm nl}^{\max} = 60 \pm 10$ with the 
variation being due to the dependence of the free-streaming
scale on the WIMP properties.

The redshift of formation of the very first WIMPy halos is very
different from the redshift $z_{\rm nl}^{\rm max}$ when hierarchical
structure formation starts at a typical place in the universe. In
order to make quantitative statements about the first WIMPy halos we
need to specify the statistical distribution of density
fluctuations. Here we assume that they have a normal (gaussian)
distribution, which is justified by two physical arguments. Firstly,
we are looking at the very first non-linear objects to form, thus
there was no non-linear physics before the rare fluctuations
that we are going to discuss enter the non-linear regime. This means
that, if the primordial fluctuations are gaussian, the fluctuations
from which the first non-linear (rare) objects form should also be close to
gaussian.  On top of that argument, the central limit theorem (in the
limit of large numbers of distributions -- one for each trial volume
-- the cumulative distribution becomes normal) justifies the use of a
gaussian distribution. A typical comoving volume of interest is the
mass collection volume of the Milky Way, which is about $1$ Mpc$^3$. As shown
above, the cut-off due to free-streaming implies that the first WIMPy
halos have a mass collecting volume of about $1$ pc$^3$. Thus we are
talking about a sample of $\sim 10^{18}$ primary halos.

As a consequence the probability that any one of the $10^{18}$ primary regions
within a comoving Milky Way volume has a density that exceeds $N\sigma$
is given by 
\begin{equation}
P(\Delta_{\rm m}({\bf x}) > N \sigma ) = 
\frac 12 \left[1-\mbox{erf}(N/\sqrt{2})\right] \,. 
\end{equation}
Such a primary halo with $\Delta_{\rm m} = N \sigma$ goes non-linear
at 
\begin{equation}
N \sigma(R_{\rm min}, z_{\rm nl}^{\max}(N)) = 1 \,. 
\end{equation} 
As fluctuations grow linear with the scale factor during the matter
dominated epoch, this condition provides us with the simple result
$z^{\rm max}_{\rm nl}(N) \approx N z_{\rm nl}^{\rm max}(N=1) \approx 
(60 \pm 20)N$. Here we consider the uncertainty in the primordial power
spectrum and in the WIMP physics. The comoving cut-off scale 
$R_{\rm min}$ is independent of the redshift $z_{\rm nl}$. 

We estimate the size and mass of the first generation of subhalos that
form at $z_{\rm nl}^{\max}$, as well as the size and mass of the very
first WIMPy halos, using the spherical collapse model (see
e.g. Reference \cite{pad}). We should caution that this simplified
model has not yet been validated in this regime where the scale
dependence of the (processed) power spectrum is relatively weak.
The mean CDM mass within a sphere of
comoving radius R is $M(R)= 1.6 \times 10^{-7} M_{\odot}(\omega_{\rm
m}/0.14) (R/{\rm pc})^3$. CDM overdensities that go non-linear have
mass twice this value i.e. roughly equal to the mass of Mars. These
WIMP halos are however much less compact than Mars. The physical size
of the first halos at turn-around (when their evolution decouples
from the cosmic expansion) is $r = 1.05 R/[1 + z_{\rm nl}^{\max}(N)]$,
which is $\sim (0.02/N)$ pc for $R_{\rm min} = 1$ pc. The first halos then 
undergo violent relaxation, decreasing in radius by a factor of two so that 
their present day radius would be of order tens of milli-pc ($N=1$), 
comparable to the size of the solar system, and smaller. 

\begin{figure}
\begin{center}
\epsfxsize=6.in
\epsfbox{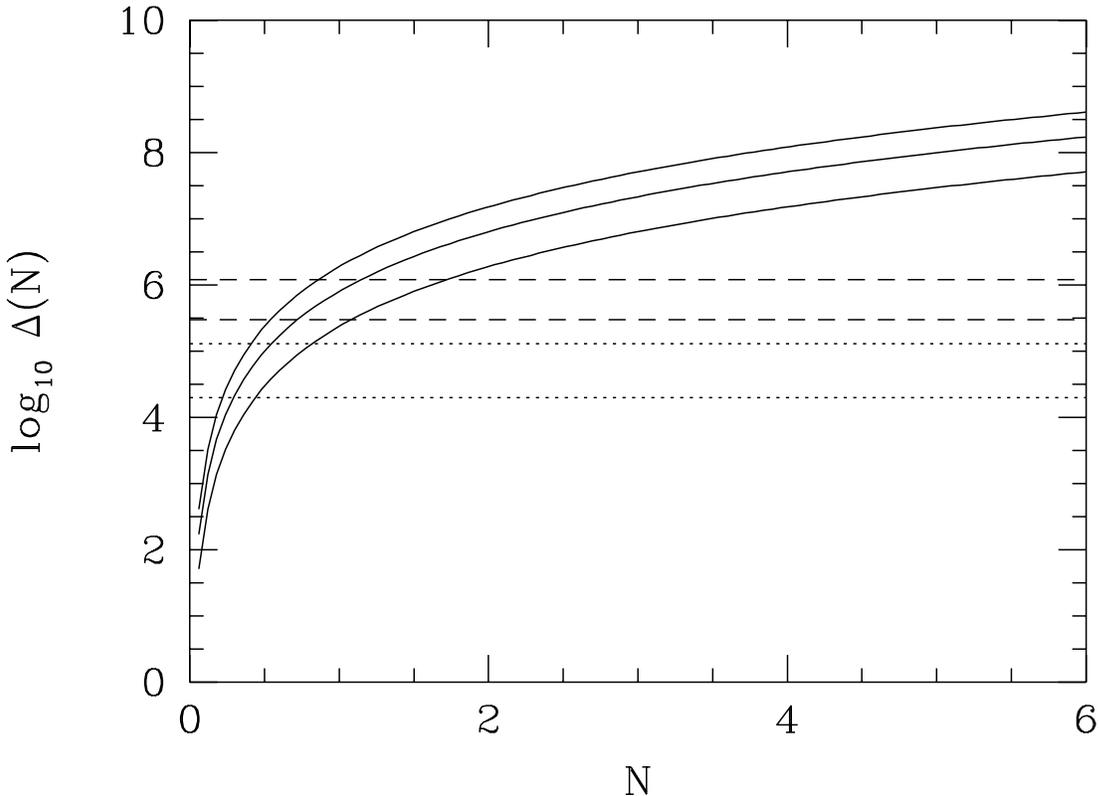}
\end{center}
\caption{The present day (mean) density contrast of $N \sigma$ fluctuations 
$\Delta(N)$ for, from top to bottom, $z_{\rm nl}^{\rm max}(N=1) = 80, 60, 40$. 
The dashed (dotted) lines indicate the range of values for the density 
contrast of the Milky Way disc (halo) in the solar neighbourhood. The
normalisation is fixed by $\sigma_8 = 0.9$. 
\label{deltaN}}
\end{figure}

A rough estimate of the relevance and chances of survival of the very first 
WIMPy halos can be made using the present day density contrast of 
these objects in the spherical collapse model. We find 
\begin{equation}
\Delta = {2 M(R_{\rm min}) \over \frac{4\pi}{3} \left[\frac{r(R_{\rm
min}, z_{\rm nl}^{\rm max}(N))}{2}\right]^3 
          \epsilon_{{\rm c}0}} = 3.7 (60 \pm 20)^3 N^3  \,. 
\end{equation}
Remarkably, the result is independent of the comoving size $R_{\rm
min}$.  As larger halos form later, their density contrast is smaller,
as $\Delta \propto z_{\rm nl}^3$.  In figure \ref{deltaN} we plot 
$\Delta(N)$ for $z_{\rm nl}^{\rm max} = 40, 60$ and $80$ and compare
it to the density contrast in the galactic disc and in the halo in the
solar neighborhood, $\Delta_{\rm disc} = (0.3 \mbox{ to } 1.2) 10^6$
and $\Delta_{\rm halo} = (0.2 \mbox{ to } 1.3) 10^5$~\cite{pdg}. For
the typical fluctuations ($N=1$) we find $\Delta = (0.2 \mbox{ to }
1.8) 10^6$, which is of the same order of magnitude as the local
density contrast of the disc.  This suggests that the typical first
WIMPy halos may not survive the highly non-linear processing which 
occurs during structure formation.

The situation is very different if we look at the rare fluctuations
($N > 1$), which are characterised by the fact that they go nonlinear
much earlier and are thus much denser than other close-by
structures. As can be seen in figure \ref{deltaN}, e.g.~$N = 3$
overdense regions lead to a range of $\Delta(N=3) = (0.6 \mbox{ to }
4.9) 10^7$, more than an order of magnitude denser than the local disc
and more than two orders of magnitude above the local halo
density. Statistically it needs $\sim 740$ comoving pc$^3$ volumes to
find one $N=3$ fluctuation. The Milky Way has a (comoving) mass
collecting radius of $\sim 1 {\rm Mpc}$ therefore, if these very first
WIMP halos survive, there would be roughly $10^{15} (10^{11})$ `rare
$N=3(6)$' WIMPy subhalos within the Milky Way.  Assuming the Milky Way
has a volume of the order of ($100$ kpc)$^3$, there would be, on
average, one `rare' $N=3(6)$ subhalo in each pc$^3$ [($20$ pc)$^3$]
volume.

What are the possible implications for direct and indirect dark matter
searches? As we discussed above, the comparison of the local halo and
disc density contrasts with those expected for the first typical
fluctuations indicates that most of the first generation of halos will
likely be destroyed during the structure formation process. In this
case, apart from within the few rare surviving subhalos, the direct
detection event rate will be only slightly lower than found using the
standard approach, which assumes a smooth dark matter distribution.
The consequences for indirect detection could be much more
dramatic. As the indirect detection rate scales with the square of the
density contrast, the `rare' subhalos discussed above could provide
`bright' point sources in the solar neighborhood, which eventually
could dominate other sources (e.g.~the center of the Milky Way). A
detailed investigation of the effect on the expected direct and
indirect dark matter search rates, as well as a detailed investigation
of the survival probability of rare fluctuations, is beyond the scope
of this paper.

\section{Discussion} 

Extensive experimental efforts are being devoted to detecting WIMPs, either 
directly in the lab or indirectly via their annihilation products. Results 
from these searches are usually quoted in terms of limits on 
the relevant WIMP interaction cross section. The expected event rates in
fact depend, in some cases very sensitively, on the dark matter
distribution and hence the formation history of the Milky Way.

Collisional damping and free-streaming produce a cut-off in the power
spectrum at a co-moving scale around $1$ pc and set the scale of the
first, smallest WIMP halos to from. In this paper we have calculated
the damping processes for generic WIMPs and examined the effect of
scale dependence of the primordial perturbation spectrum. We have fixed 
the parameters of the $\Lambda$CDM model, $\omega_{\rm m}, \omega_{\rm b},
\Omega_\Lambda$ and $A$ (resp.~$\sigma_8$), to the best-fit WMAP values.
A discussion of the effect of varying CDM density has been provided in
our previous work \cite{ghs}. 

We have found that the smallest scale fluctuations go non-linear (more
precisely typical one-sigma fluctuations collapse to from dark matter
halos) at a red-shift in the range 40-80, with the first WIMP halos
forming significantly earlier from rare large fluctuations. Finally we
estimated the properties of both the typical and first small halos to
form using the spherical collapse model.  The mass of the halos is
independent of the size of the fluctuation from which they form,
however the first rare fluctuations to collapse form more compact,
denser halos.

Neglecting collisional damping and free streaming of WIMPs
would result in monotonically increasing power of density fluctuations
on small scales. As a consequence, there would be a divergence of the 
energy density of the fluctuations at small scales and some kind of 
regularization procedure would be required 
to make the hierarchical picture of structure formation well-defined. 
The collisional damping and free streaming of WIMPs 
regularize the power spectrum by providing a physical cut-off scale. 

Numerical studies of the formation of the first WIMPy halos
have recently been carried out by Diemand et al.~\cite{dms}. 
The subsequent evolution of these halos, and the resulting present
day dark matter distribution remains an important outstanding issue.

\ack

We would like to thank Lars Bergstr\"om, Joakim Edsj\"o, Ariel Goobar,
Abraham Loeb, Ben Moore, Kerstin Paech, Mia Schelke, James Taylor, 
Tom Theuns, Licia Verde and Matias Zaldarriaga for valuable discussions.  
AMG was supported by the Swedish Research Council and PPARC. 
SH was supported by the Wenner-Gren Foundation.

\end{document}